\documentclass[prd,aps,twocolumn,a4paper,floatfix,nofootinbib]{revtex4-2}

\usepackage[utf8]{inputenc}
\usepackage{graphicx,psfrag}
\usepackage{mathrsfs}
\usepackage{amsmath,amsfonts,amssymb}
\usepackage{multirow}
\usepackage{diagbox}
\usepackage{comment}
\usepackage{xcolor}
\usepackage{enumerate}
\usepackage{booktabs}
\usepackage[normalem]{ulem}
\usepackage{hyperref}
\usepackage{mathtools}
\usepackage{siunitx}
\usepackage{textgreek}
\hypersetup{
    colorlinks = true,
    linkcolor = {blue},
    citecolor = {blue},
    urlcolor = {blue},
    linkbordercolor = {white},
    }

\usepackage{color}
\definecolor{cyan}{rgb}{0,0.9,0.9}
\definecolor{orange}{rgb}{0.9,0.5,0}
\definecolor{magenta}{rgb}{1,0,1}
\definecolor{purple}{rgb}{0.8,0.4,0.8}
\definecolor{gray}{rgb}{0.8242,0.8242,0.8242}
\definecolor{green}{rgb}{0.,0.8,0.}

\def\bam{{\textsc{bam}}}
\def\sgrid{{\textsc{sgrid}}}

\usepackage[normalem]{ulem}

\begin{document}

\title{Magneto-Hydrodynamic Simulations of Eccentric Binary Neutron Star Mergers}

\author{Anna \surname{Neuweiler}$^{1}$}
\author{Tim \surname{Dietrich}$^{1,2}$}
\author{Bernd \surname{Br\"ugmann}$^{3}$}

\affiliation{${}^1$ Institut f\"ur Physik und Astronomie, Universit\"at Potsdam, Haus 28, Karl-Liebknecht-Str. 24/25, 14476, Potsdam, Germany}
\affiliation{${}^2$ Max Planck Institute for Gravitational Physics (Albert Einstein Institute), Am M\"uhlenberg 1, Potsdam 14476, Germany}
\affiliation{${}^3$ Theoretical Physics Institute, University of Jena, 07743 Jena, Germany}

\date{\today}

\begin{abstract}
Highly eccentric binary neutron star mergers exhibit unique dynamical and observational signatures compared to quasi-circular ones in terms of their gravitational wave signal and the ejection of matter, leading to different electromagnetic counterparts. 
In this article, we present general relativistic magneto-hydrodynamic simulations of binary neutron star systems on highly eccentric orbits. While in quasi-circular binaries, the influence of the magnetic field is too weak to affect the general pre-merger dynamics, the close encounters in eccentric systems could potentially trigger magneto-hydrodynamic instabilities. Therefore, we investigate possible effects before, during, and after the merger for a total of three different systems with varying initial eccentricity.

We study the $f$-mode oscillations excited by tidal interaction in close encounters and find good agreement with predicted $f$-mode frequency estimates. However, our simulations reveal no significant differences compared to results neglecting the magnetic field. Although we observe a rearrangement of the poloidal structure of the magnetic field inside the stars, there is no relevant increase in the magnetic energy during the encounters. Also, during the merger, the amplification of the magnetic field seems to be largely independent of the eccentricity in our systems. Consistent with studies of merging non-magnetized binary neutron stars, we find a correlation between eccentricity and mass ejection, with a higher impact parameter leading to a larger amount of unbound material. 
\end{abstract}

\maketitle

\section{Introduction}
\label{sec:Intro}
Due to their rich phenomenology, binary neutron star (BNS) mergers are among modern astrophysics' most fascinating research topics. In particular, since the first detection of gravitational waves (GWs) from a neutron star merger in August 2017, GW170817~\cite{LIGOScientific:2017vwq}, the research field on BNS systems has received increasing attention. Along the GW signal, electromagnetic (EM) counterparts, the gamma-ray burst (GRB) GRB170817A~\citep{Goldstein:2017mmi, Savchenko:2017ffs} and the kilonova AT2017gfo~\citep{LIGOScientific:2017pwl}, were observed associated with the same source. Among others, this multi-messenger event allowed us to set tighter constraints on the equation of state (EOS) for neutron stars (NSs), e.g.,~\cite{Radice:2016rys,Margalit:2017dij,LIGOScientific:2018cki,Pang:2022rzc}, provided a new method to measure the cosmic expansion, e.g.,~\cite{LIGOScientific:2017adf,LIGOScientific:2019zcs,Hotokezaka:2018dfi,LIGOScientific:2018gmd,Dietrich:2020efo}, and established BNS merger as sources for GRBs, e.g., \cite{Troja:2017nqp,Hallinan:2017woc,Mooley:2017enz,Lazzati:2017zsj}, and sites for heavy element formation via the $r$-process, e.g.,~\cite{Lattimer:1974slx,Rosswog:1998hy,Korobkin:2012uy,Wanajo:2014wha}. 

The correct interpretation of the observed data relies on precise theoretical models for GWs and EM signals. Due to the high complexity of the system, accurate predictions of the merger dynamics require numerical-relativity (NR) simulations of BNS mergers that solve Einstein's field equations. Moreover, the development of models that can be used for parameter estimation depends on an extensive coverage of the entire parameter space for BNS systems. Current NR simulations span a significant portion of NS masses, mass-ratios, and spins for different EOSs to describe the NS interior \cite{Dietrich:2018phi,Kiuchi:2017pte,Foucart:2018lhe,Kiuchi:2019kzt,Boyle:2019kee,Gonzalez:2022mgo}. While most BNS simulations are performed for quasi-circular systems, only a few studies have analyzed eccentric BNS systems with significant eccentricities at merger in full general relativity (GR), e.g., \cite{Gold:2011df,East:2012ww,Radice:2016dwd,Papenfort:2018bjk,Chaurasia:2018zhg}. 

With the emission of gravitational radiation, the eccentricity decays, and the orbit is typically expected to be nearly circular before reaching frequencies observable by current GW detectors. However, some formation channels predict non-negligible eccentricity for BNS mergers in dense stellar environments during dynamical encounters, e.g., through tidal captures, binary-single star or binary-binary interactions. Specifically, the eccentricity of a BNS system can be enhanced by the exchange of angular momentum via the so-called Lidov-Kozai mechanism~\cite{Kozai:1962,Lidov:1962}. Thus, the detection of a non-zero eccentric compact binary merger would be a decisive indication of a dynamical formation channel in stellar clusters or triple systems. Despite the lack of a comprehensive eccentric BNS GW models, several searches for eccentric binary systems have been performed in current GW catalogs, e.g., \cite{LIGOScientific:2019dag,Cheeseboro:2021rey,Pal:2023dyg,LIGOScientific:2023lpe,Ravichandran:2023qma,Dhurkunde:2023qoe,Gadre:2024ndy}, highlighting the growing interest. Recently, \cite{Morras:2025xfu} also found strong indications of non-zero eccentricity in the black hole-neutron star merger GW200105 indicating a dynamical formation channel. For completeness, we also refer to several works extending binary black hole models to the eccentric parameter space, e.g., \cite{Habib:2019cui,Huerta:2019oxn,Ramos-Buades:2019uvh,Healy:2022wdn,Ramos-Buades:2022lgf,Wang:2023vka,Wang:2024jro,Nee:2025zdy}.

Gravitational waveforms of highly eccentric systems differ significantly from the classical chirp signal of quasi-circular inspirals. Instead of a steadily increasing amplitude and frequency, eccentric systems exhibit bursts of GW radiation at each close encounter. This phenomenon has been studied first using Newtonian orbits together with leading-order relativistic expressions for the radiation and the evolution of the orbital parameters~\cite{Peters:1963ux,Turner:1977}. Later, numerical simulations exploring eccentric BNS systems were performed in Newtonian gravity, e.g., \cite{Lee:2010,Rosswog:2012wb} and also in full GR, e.g., \cite{Gold:2011df,East:2012ww,Radice:2016dwd,Papenfort:2018bjk,Chaurasia:2018zhg}. These studies have shown that tidal interactions during the close, pre-merger encounters of eccentric BNS systems can excite $f$-mode oscillations of the stars, generating a characteristic GW signal. These oscillations can be used to study dynamical tides, providing valuable insights into NS seismology and EOS properties \cite{Takatsy:2024sin}. 
Furthermore, the mass of the remnant disk and the amount of unbound matter vary depending on the impact parameter. Thus, eccentric mergers can produce significantly more ejecta, leading to brighter EM counterparts \cite{East:2011xa,Rosswog:2012wb,Radice:2016dwd,Papenfort:2018bjk}.

However, none of the studies above consider effects of magnetic fields, and the results are limited to simulations of general-relativistic hydrodynamics \cite{Gold:2011df,East:2012ww,Chaurasia:2018zhg} with a leakage scheme to model neutrino emission in \cite{Radice:2016dwd,Papenfort:2018bjk}. Magneto-hydrodynamic effects typically influence the dynamics primarily after the merger. Due to the Kelvin-Helmholz instability (KHI), e.g.,~\cite{Kiuchi:2015sga, Kiuchi:2017zzg,Giacomazzo:2014qba}, the Rayleigh-Taylor instability, e.g.,~\cite{Skoutnev:2021chg}, and the magneto-rotational instability (MRI), e.g.,~\cite{Siegel:2013nrw,Kiuchi:2023obe}, magnetic fields amplify at merger and can alter the lifetime of the merger remnant, shape the outflow of neutron-rich matter, and possibly trigger a relativistic jet. In quasi-circular binaries, the effect of the magnetic field is too weak to influence the dynamics before merger. Nevertheless, as pointed out in \cite{Radice:2016dwd}, the pre-merger encounters of eccentric BNS systems could trigger magneto-hydrodynamic instabilities, and the magnetic fields could become dynamically relevant even before the merger.

In this work, we study the potential effects of the magnetic field in eccentric BNS mergers. With our recent extension towards general-relativistic magneto-hydrodynamics (GRMHD) simulations~\cite{Neuweiler:2024jae} with our NR code \bam~\cite{Bruegmann:2006ulg,Thierfelder:2011yi}, we perform, up to our knowledge, the first GRMHD simulations of highly eccentric BNS systems. The goal is to gain a comprehensive understanding of how the inclusion of the magnetic field changes the dynamics of the system, possibly before but also after the merger. To achieve this, we consider a small set of equal-mass BNS systems at a fixed initial distance of $\sim 355.6$~km and vary the initial eccentricity. We investigate possible changes in the magnetic field during the close encounters before the merger. In addition, simulations without a magnetic field are performed to analyze the effects on the GW signal, the outflow properties, and the ejecta mass as well as the remnant system.

The article is structured as follows: 
Section~\ref{sec:Methods} provides details on the configurations and methods used, including the construction of initial data and the dynamical evolution of the systems. In Sec.~\ref{sec:Results}, we present the simulation results and analyze effects of the magnetic field before, during, and after the merger. Finally, we conclude in Sec.~\ref{sec:Conclusions} and summarize our main findings. In this article, we apply a metric with $(-,+,+,+)$ signature and geometric units with $G=c=M_\odot=1$, unless otherwise specified. 

\section{Setups and Numerical Methods}
\label{sec:Methods}

\subsection{Initial Data Construction} 

We construct the initial data for the eccentric BNS systems with the pseudo-spectral code \sgrid~\cite{Tichy:2009yr,Tichy:2012rp,Dietrich:2015pxa,Tichy:2019ouu}. In order to solve the Einstein Constraint Equations, \sgrid\ uses the extended conformal thin sandwich approach~\cite{York:1998hy,Pfeiffer:2002iy}. We apply the method introduced in \cite{Moldenhauer:2014yaa,Dietrich:2015pxa} to construct eccentric BNS systems. For completeness, we briefly outline here the approach. 
We assume the stars start at apoapsis and define the centers of the stars $A$ and $B$ as
\begin{align}
    x_A &=  a \left(1+e\right) \frac{M^B}{M^A + M^B} + x_{\rm cm}, \\
    x_B &= -a \left(1+e\right) \frac{M^A}{M^A + M^B} + x_{\rm cm}.
\end{align}
Here, $x_{\rm cm}$ denotes the center of mass of the full binary system. The semimajor axis $a$, the semiminor axis $b$, and the eccentricity $e$ with $e^2=\left(1-b^2/a^2\right)$ describe the elliptic orbit. We can approximate the actual motion of the stars by inscribed circles into the elliptical orbit with centers at $x_{c_{A,B}} = x_{\rm cm} + e\left(x_{A,B} - x_{\rm cm}\right) $. This helps us to assume an approximate `helliptical' (combination of helical and elliptical motion) Killing vector
\begin{equation}
    k^\alpha_{A,B} = t^\alpha + \omega \left[ \left(x-x_{c_{A,B}}\right)y^\alpha - y x^\alpha \right] +\frac{v_r}{r_{AB}}r^\alpha,
\end{equation}
with the vectors ${\bf t} = \partial_t$, ${\bf x} = \partial_x$, and ${\bf y} = \partial_y$ generating the translations in the $t$, $x$, and $y$ directions. $r^\alpha = \left(0,x,y,z\right)$ points in the radial direction and $r_{AB}$ is the distance between the two star centers. The parameter $v_r$ defines the radial velocity and $\omega$ the angular velocity. For this work, we set $v_r = 0$ and vary the input parameter $e$ to construct systems with different eccentricities.
Although this method neglects oscillations of the stars that may have been induced by previous close encounters, we assume such initial oscillations to be small and thus their absence not to affect later oscillations significantly.

\subsection{Dynamical Evolution}

For the evolution of the BNS systems, we use the \bam\ code~\cite{Bruegmann:2006ulg,Thierfelder:2011yi,Dietrich:2015iva,Bernuzzi:2016pie} with our new GRMHD implementation \cite{Neuweiler:2024jae}. \bam\ solves the Einstein's field equations performing a $3+1$ decomposition. To evolve spacetime, we use the Z4c reformulation with constraint damping terms \cite{Bernuzzi:2009ex,Hilditch:2012fp} combined with the 1+log slicing~\cite{Bona:1994a} and gamma-driver shift~\cite{Alcubierre:2002kk} conditions.

Our GRMHD implementation follows the Valencia formulation~\cite{Font:2008fka,Marti:1991wi,Banyuls:1997zz,Anton:2005gi}. Here, we apply the ideal GRMHD approximation, which assumes infinite conductivity and zero resistivity.\footnote{While we assume that ideal GRMHD is a fairly accurate description for most magneto-hydrodynamical processes in BNS mergers, this assumption breaks when modeling the magnetosphere outside the NS. Therefore, in the following analysis for pre-merger encounters, we neglect the magnetosphere and focus on the magnetic field inside the stars. However, we point towards studies using force-free GRMHD, e.g, \cite{Most:2020ami,Skiathas:2025pnj}, for a discussion of possible pre-merger radio bursts due to interacting magnetospheres.}
In order to ensure that the divergence-free constraint for the magnetic field remains satisfied and no unphysical magnetic monopoles form, we use the hyperbolic \textit{divergence cleaning} scheme from \cite{Liebling:2010bn,Mosta:2013gwu}. For more details on the implementation and a discussion on the performance compared to other ideal GRMHD codes, we refer the reader to \cite{Neuweiler:2024jae}. We use in this work a piecewise polytropic fit of the SLy EOS~\citep{Douchin:2001sv} following \cite{Read:2008iy} to describe the interior of the NS using four pieces (three for the core of the NS and one for its crust). Thermal effects are included by extending the zero-temperature EOS by a thermal pressure $P_{\rm th} = \left(\Gamma_{\rm th} - 1\right) \rho \epsilon_{\rm th}$ \cite{Bauswein:2010dn}, where $\epsilon_{\rm th}$ is the thermal part of the specific internal energy. We set $\Gamma_{\rm th} = 1.75$.

The simulation grid consists of a hierarchy of nested refinement levels labeled by $l=0,1,...,L-1$. Each of the $L$ levels contains one or more Cartesian boxes with fixed grid spacing. Following a $2:1$ refinement strategy, the grid spacing level on $l$ is given by $h_l = h_0 / 2^l$. The Cartesian boxes for the inner refinement levels with $l \geq l_{\rm mv}$ can dynamically move, tracking the compact objects during the evolution. Thereby, each box has $n$ or $n_{\rm mv}$ grid points per direction for outer, non-moving or inner, moving refinement levels, respectively.

For the time integration, we use a fourth-order Runge-Kutta scheme and a Courant-Friedrichs-Lewy coefficient of 0.25. \bam\ employs the Berger-Oliger scheme~\citep{Berger:1984zza} for local-time stepping (see \cite{Bruegmann:2006ulg}) and the Berger-Colella scheme~\citep{Berger:1989a} to ensure conservation of baryonic mass, energy, and momentum across refinement boundaries (see \cite{Dietrich:2015iva}). While the spatial derivatives of the metric are approximated by fourth-order finite difference stencils, the fluid variables are modeled using high-resolution shock-capturing techniques to calculate numerical fluxes between the grid cells. In this work, we use the Harten, Lax, and van Leer (HLL) Riemann solver \cite{Harten:1983} with a two-wave approximation. For the reconstruction of the field variables to the individual grid cell interfaces, the fifth-order weighted-essentially-non-oscillatory (WENOZ) scheme~\cite{Borges:2008} is applied. In low density regimes with $\rho < 10^{-7} \rho_c$ with $\rho_c$ being the central density of the initial stars, we apply the procedure outlined in \cite{Neuweiler:2024jae} to avoid enhanced oscillations and ensure the physical validity of reconstructed variables. In particular, we examine the oscillation of all variables and fall back to a lower-order scheme if necessary, in this case a third-order convex-essential-non-oscillating (CENO3) scheme~\cite{Liu:1998,Zanna:2002qr}. Finally, we verify the physical validity of the reconstructed values by demanding a positive rest-mass density and a positive pressure. If this is not given even with the lower-order scheme, we resort to a linear reconstruction method.

\begin{table}[t!]
    \centering
    \caption{BNS configurations. From left to right, we list: The configuration name, the input eccentricity $e$, the 3PN eccentricity estimate $\hat{e}_{3 \rm PN}$ following \cite{Mora:2003wt}, the individual gravitational masses $M^{\rm A,B}$, the individual baryonic masses  $M^{\rm A,B}_{\rm b}$, the initial GW frequency $M \omega^0_{22}$, the ADM mass $M_{\rm ADM}$, and the ADM angular momentum $J_{\rm ADM}$.}
    \label{tab:configurations}
    \begin{tabular}{l||cc|cc|ccc}
    \toprule 
    name & $e$ & $\hat{e}_{3 \rm PN}$ & $M^{\rm A,B}$ & $M^{\rm A,B}_{\rm b}$ & $M \omega^0_{22}$ & $M_{\rm ADM}$ & $J_{\rm ADM}$ \\ 
    \hline
    \hline
    ecc0.50 & 0.50 & 0.54 & 1.35 & 1.4946 & 0.0107 & 2.6875 & 8.4635 \\ 
    ecc0.55 & 0.55 & 0.59 & 1.35 & 1.4946 & 0.0113 & 2.6871 & 8.0206 \\ 
    ecc0.60 & 0.60 & 0.64 & 1.35 & 1.4946 & 0.0120 & 2.6867 & 7.5531 \\ 
    \bottomrule
    \end{tabular}
\end{table}

\subsection{Configurations}

We study three different physical configurations with varying input eccentricities $e$ of $0.5$, $0.55$, and $0.6$. The configurations are summarized in Tab.~\ref{tab:configurations}. For this work, we focus on non-spinning, equal-mass BNS systems with baryonic masses of $M^A_b = M^B_b \backsimeq 1.4946\ M_\odot$ with fixed initial separation of $\sim 355.6\ \rm km$. The input eccentricities and parameters are chosen to be comparable to those used in \cite{Chaurasia:2018zhg}.

\begin{figure*}[htp!]
    \centering
    \includegraphics[width=\linewidth]{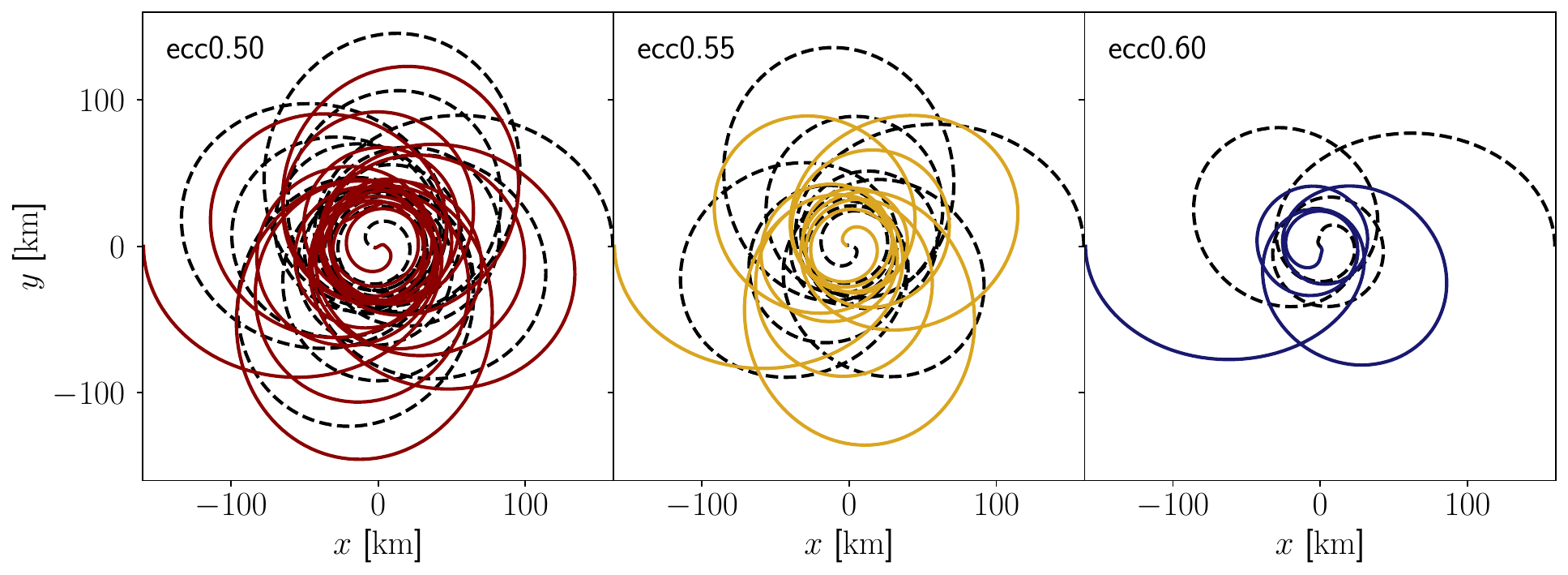}
    \caption{Orbital trajectories shown via colored solid lines and black dashed lines for the two NS centers for the simulations at R2 resolution with different input eccentricities: $e=0.50$ (left panel), $e=0.55$ (middle panel) and $e=0.60$ (right panel).}
    \label{fig:Inspiral}
\end{figure*}

We list in Table~\ref{tab:configurations} additionally to the input eccentricities a post-Newtonian (PN) estimate from the Arnowitt-Deser-Misner (ADM) expressions for the energy and angular momentum as in \cite{Dietrich:2015pxa,Chaurasia:2018zhg}. We use the 3PN expression computed following \cite{Mora:2003wt} and define:
\begin{align}
    \hat{e}^2_{3PN} =& 1 - 2 \xi + \left[-4 -2 \nu + \left(-1 +3 \nu \right) \zeta \right] E_b \nonumber \\
    &+\left[ \frac{20 -23 \nu}{\xi} - 22 + 60 \nu + 3 \nu^3 - \left(31 \nu+4 \nu^2 \right) \xi \right] E_b^2 \nonumber \\
    &+\left[\frac{-2016 + \left(5644 -133 \pi^2 \right) \nu - 252 \nu^2}{12 \xi^2} \right. \nonumber \\
    &+      \frac{4848 +\left(-21128 +369 \pi^2\right) \nu + 2988 \nu^2}{24 \xi} \nonumber \\ 
    & -20 +298 \nu-186 \nu^2 - 4 \nu^3 \nonumber \\
    & \left. + \left(-30 \nu+\frac{283}{4} \nu^2 + 5 \nu^3 \right) \xi \right] E_b^3 ,
\end{align}
with $\xi = -E_bl^2$ and the symmetric mass ratio $\nu = M^A M^B / M^2$. Here, $E_b = \left(M_{\rm ADM}/M-1\right)/\nu$ is the binaries reduced binding energy and $l = \left(J_{\rm ADM} - S^A - S^B\right)/\left(M^2 \nu\right)$ is the specific orbital angular momentum. 

The magnetic field is initialized assuming a purely poloidal field inside each star defined by a vector potential with:
\begin{align}
    A_x &= -\left(y - y_{\rm NS}\right) A_b \max(p - p_{\rm cut},0)^2, \\
    A_y &=  \left(x - x_{\rm NS}\right) A_b \max(p - p_{\rm cut},0)^2, \\
    A_z &=  0,
\end{align}
where $x_{\rm NS}$ and $y_{\rm NS}$ refer to the coordinate centre of each NS. For the simulations performed, we set $A_b = 1000$ to obtain a maximum magnetic field strength of the order of $10^{15}$\,G inside the stars. The cut-off pressure is set to $p_{\rm cut}=0.004 \times p_{\rm max}$, where $p_{\rm max}$ is the initial maximum pressure in the NS at $t=0$\,ms. 

\begin{table}[t!]
    \centering
    \caption{Grid configurations. From left to right, we list: The resolution name, the number of points in the non-moving refinement boxes $n$, the number of points in the moving refinement boxes $n_{\rm mv}$, the number of refinement levels $L$, the level $l_{\rm mv}$ for which $l>l_{\rm mv}$ can dynamically move during the evolution, the grid spacing on the coarsest level $h_0$, and the grid spacing on the finest level $h_L$. Note that we employ reflection symmetry across the orbital plane, which gives us the division by $2$ for the grid points.}
    \label{tab:simulation_grid}
    \begin{tabular}{l||cc|cc|cc}
    \toprule
    res. & $n$ & $n_{\rm mv}$ & $L$ & $l_{\rm mv}$ & $h_0$ & $h_L$ \\ 
    \hline
    \hline
    R1 & $256^3/2$ & $128^3/2$ & $8$ & $2$ & $30$ & $0.1171875$ \\ 
    R2 & $384^3/2$ & $192^3/2$ & $8$ & $2$ & $20$ & $0.078125$ \\ 
    \bottomrule
    \end{tabular}
\end{table}

We run each system for two different resolutions R1 and R2 as indicated in Table~\ref{tab:simulation_grid}. Additionally, we run simulations with pure general-relativistic hydrodynamics at the lower resolution R1 to evaluate the differences between the results with and without magnetic field.

\section{Results}
\label{sec:Results}

We perform a comprehensive analysis of the simulations by investigating potential effects of the magnetic field before, during, and after the merger. On the one hand, we compare results for different eccentricities, i.e., how previous encounters and different impact parameters at merger change the magnetic field structure. On the other hand, we evaluate how the inclusion of a non-zero magnetic field affects simulation results and dynamics of eccentric BNS systems.

\subsection{Encounters Before the Merger}
\label{sucsec:PreMerger}

\subsubsection{Qualitative Discussion of the Inspiral}
\label{subsubsec:QualitativeDiscussion}

Keeping the initial separation fixed for all configurations and only varying the input eccentricity, the number of orbits until the merger decreases with increasing $e$. Thus, the system with $e=0.50$ merges after $\sim 18.3$~orbits, the system with $e=0.55$ after $\sim 10.8$~orbits, and the system with $e=0.60$ after $\sim 4.2$~orbits. We show in Fig.~\ref{fig:Inspiral} the orbital tracks of the NSs for each configuration at R2 resolution. The orbits undergo apsidal precession, leading to the expected rosetta-like shape.

\begin{figure}[t!]
    \centering
    \includegraphics[width=\linewidth]{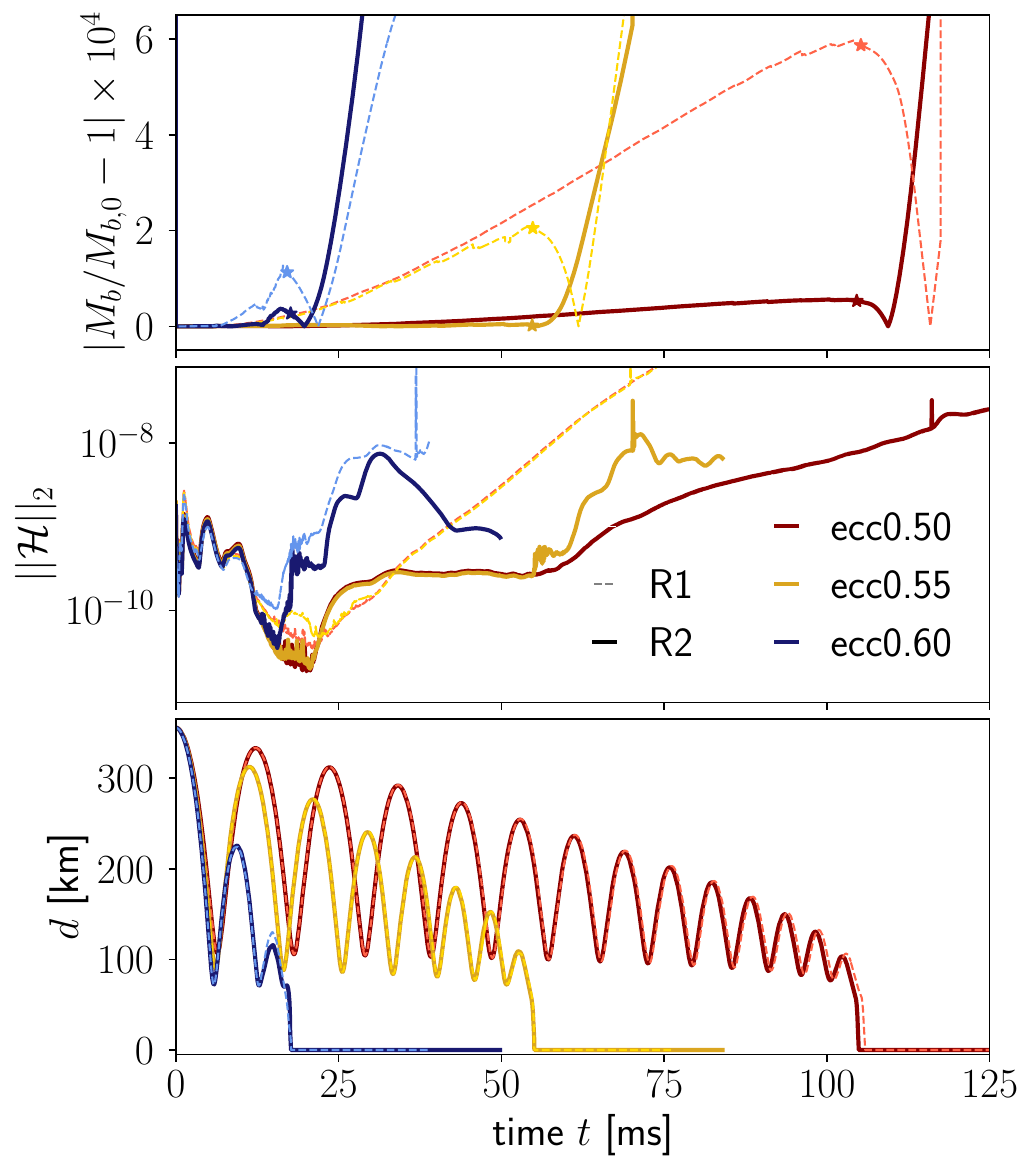}
    \caption{Time evolution of selected metrics for the eccentric BNS systems at resolution R1 (dashed lines) and R2 (solid lines). We show relative difference of the baryonic rest mass $M_b$ from its initial values $M_{b,0}$ (top panel), the $L^2$ norm of the Hamiltonian constraint $||\mathcal{H}||_2$ (middle panel), and the proper distance $d$ (bottom panel). The star marker in the top panel indicate the merger time for the each simulation. Both monitoring quantities, $M_b$ and $||\mathcal{H}||_2$, are extracted from refinement level $l=1$.}
    \label{fig:BNSmetrics}
\end{figure}

The bottom panel of Fig.~\ref{fig:BNSmetrics} shows the time evolution of the proper distance $d$ over time for each system with resolutions R1 and R2. The decrease in eccentricity is clearly visible as the proper distance at apoapsis decreases with time. This happens faster in the simulation with higher $e$. During close encounters, the individual stars deform due to the stronger tidal forces of the companion, and the stars begin to oscillate, resulting in an imprint in the GW signal (see Sec.~\ref{subsubsec:GWsignal}). 

To ensure the validity of our results discussed in the later sections of this work, we monitor the time evolution of some selected metrics in Fig.~\ref{fig:BNSmetrics}: The L2 norm of the Hamiltonian constraint $||\mathcal{H}||_2$, as well as the relative difference of the baryonic rest mass $M_b$ from its initial value $M_{b,0}$. Overall, both quantities show relatively robust behavior. $||\mathcal{H}||_2$ initially decreases due to the constraint damping properties of the Z4c evolution scheme, but increases again at time of merger. The ecc0.50 and ecc0.55 simulations at R1 resolution show an almost exponential increase after $\sim 20\ \rm ms$, which we attribute to boundary effects. These decrease at the higher R2 resolution, where $||\mathcal{H}||_2$ remains below $\sim 5 \times 10^{-8}$ during the entire simulations. The conservation of the baryonic rest-mass also improves significantly at higher resolution demonstrating the convergence of the evolution scheme. The relative difference remains below $6 \times 10^{-4}$ before the merger in each simulation. We note that after the merger, when matter is ejected, the density of the expanding ejecta decreases and a fraction of the material falls below the atmosphere threshold. As a result, the conservation of the baryonic rest mass is no longer guaranteed.

\subsubsection{Gravitational Wave Signatures and $f$-mode Oscillations}
\label{subsubsec:GWsignal}

\begin{figure}[t!]
    \centering
    \includegraphics[width=\linewidth]{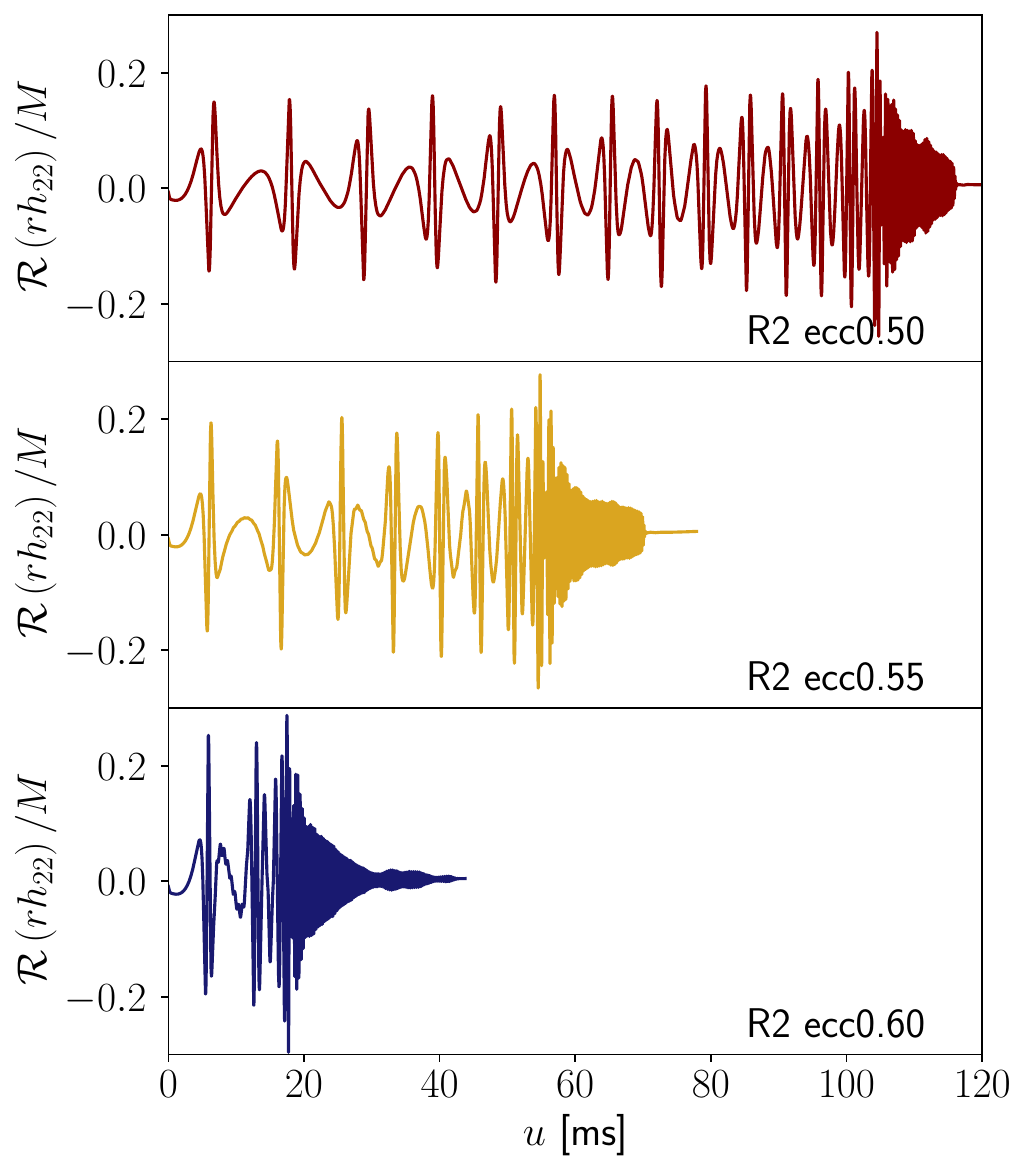}
    \caption{Real part of the $\left(2,2\right)$-mode of the GW strain, $rh_{22}$, against the retarded time $u$ for all systems at R2 resolution. The signals are extracted at $r_{\rm extr} = 1200\ M_\odot$ on refinement level $l=1$.}
    \label{fig:GWsignal}
\end{figure}

We show the $l=m=2$ modes of the GW strain in Fig.~\ref{fig:GWsignal} for all systems. The signal is extracted at the areal radius in isotropic coordinates $r_{\rm extr} = 1200\ M_\odot$ at retarded time
\begin{equation}
    u = t- r_{\rm extr} - 2M \ln{\left(\frac{r_{\rm extr}}{2M} -1 \right)},
\end{equation}
with $t$ being the simulation time and $M$ the total mass of the system. It is evident that the GW emission differs significantly from the characteristic chirp signal of quasi-circular BNS systems. Instead of a steady increase in amplitude and frequency, the signal exhibits GW bursts at each close encounter. Furthermore, there are quasi-normal mode oscillations superimposed on the GW signal of the orbital motion. Those are especially visible in the ecc0.60 simulation after the first encounter. For the ecc0.50 and ecc0.55 systems, those oscillations are significantly smaller and hardly visible in the GW strain.

\begin{figure*}[t!]
    \centering
    \includegraphics[width=\linewidth]{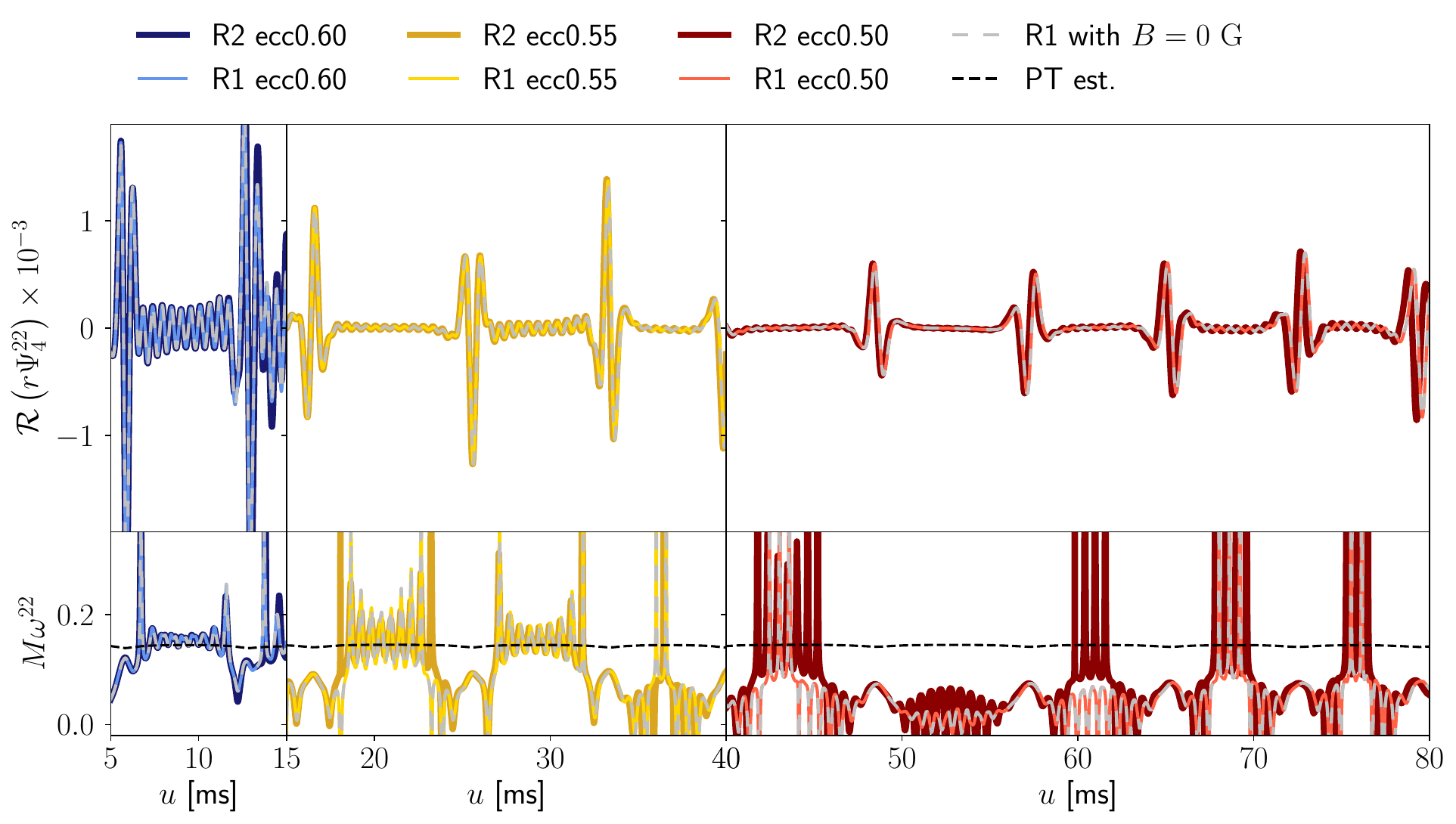}
    \caption{Real part of $r\Psi^{22}_4$ (top panels) and the corresponding GW frequency $M\omega^{22}$ (bottom panels) against the retarded time $u$. We show the results for all systems at resolutions R1 and R2 together with the results from the simulations setting $B=0\ \rm G$ at R1 resolution (grey dashed lines) as comparison. The signals are extracted at $r_{\rm extr} = 1200\ M_\odot$ on refinement level $l=1$. Additionally, we plot the PT estimate for the $f$-mode frequency as black dashed line.}
    \label{fig:Psi4}
\end{figure*}

To better assess the $f$-mode oscillations, Fig.~\ref{fig:Psi4} shows the $\left(2,2\right)$-mode of the $r\Psi_4$, for which the oscillations are easier to observe. Again, we show the results extracted at radius $r_{\rm extr} = 1200\ M_\odot$. For each system, we plot a representative time window of the inspiral for the simulations with R1 and R2 resolution as well as for the simulations without magnetic fields with R1 resolution in dashed lines. The instantaneous GW frequency $M\omega^{22}$ computed from the phase of $r\Psi^{22}_4$ is presented in the bottom panels of Fig.~\ref{fig:Psi4}.

As in Fig.~\ref{fig:GWsignal}, the GW bursts can be clearly seen during the close encounter when the amplitudes of the $r\Psi^{22}_4$ signal increase significantly. Subsequent to the bursts after the close encounter and the tidal interaction, the signal shows oscillations with lower amplitudes, which are associated with the excited oscillations of the NSs. Overall, the amplitude of the GW burst signals as well as the induced oscillations appear to be stronger for systems with higher eccentricities, i.e., for ecc0.60. 

In order to analyze the $M\omega^{22}$ oscillation frequency, we additionally compute the perturbation-theoretic (PT) estimate of the $f$-mode excitation frequency based on the $f$-love relation from \cite{Chan:2014kua}. For the SLy EOS and a gravitational mass of $1.35\ M_\odot$, we obtain an $f$-mode frequency of $1.76\ \rm kHz$. To account for the contribution of the redshift, we additionally use the leading-order PN expression $\omega^{\rm redshift}_{f-{\rm mode}} = \left(1 - M^{B}/d\right) \omega_{f-{\rm mode}}$ \cite{Krisher:1993} with the proper distance $d$ between the two stars. While the signal represents the sum of the oscillations of both stars on the extraction sphere, it is sufficient here to consider only star $A$, as our binary systems are symmetric. The estimate is shown as dashed lines in the lower panels of Fig.~\ref{fig:Psi4}.

The $M\omega^{\rm redshift}_{f-{\rm mode}}$ results agree relatively well with the $M\omega^{22}$ frequencies of our simulations. In particular, the GW frequency of ecc0.60 matches the prediction of the estimated $f$-mode frequency. For ecc0.55, the oscillation frequency shows a larger fluctuation, but the PT estimate is well within the range. In the case of the ecc0.50 system, the frequencies cover an even wider range. We assume that here the wider range is a result of induced higher-frequency overtones. In fact, the $M\omega^{\rm redshift}_{f-{\rm mode}}$ estimate lies in the lower part of the GW frequency range. We also observe that in the ecc0.55 and 0.50 simulations, where we have multiple encounters that excite $f$-modes, the oscillation is canceled after some encounters. This shows that oscillations can also be annihilated if the tidal excitation is out of phase with already existing oscillations, agreeing with results from \cite{Chaurasia:2018zhg}. 

\begin{figure*}[t!]
    \centering
    \includegraphics[width=\linewidth]{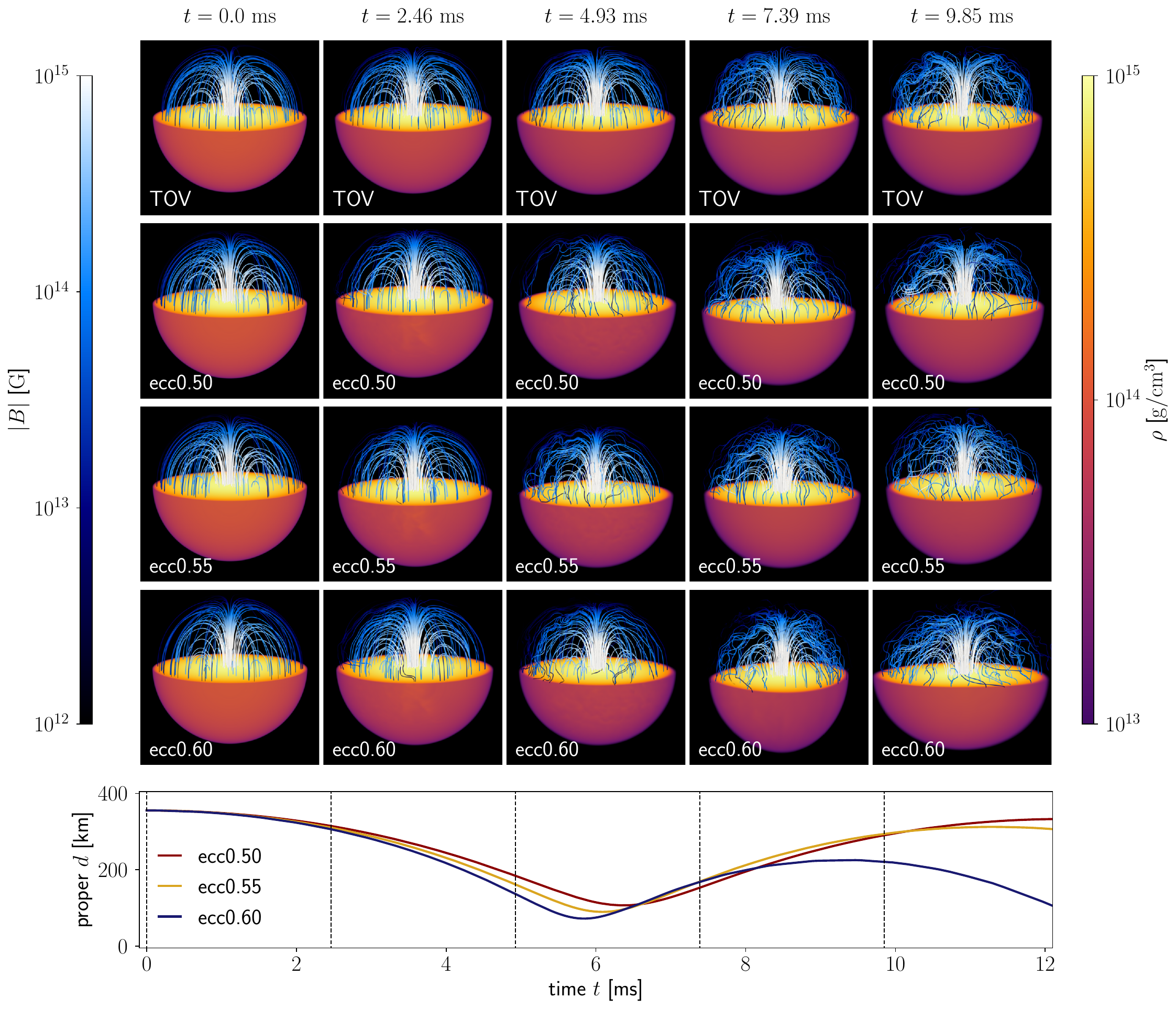}
    \caption{Three-dimensional snapshots of the magnetic field lines inside the NS at the first close encounter. For each panel, the lower half shows the rest-mass density $\rho$ with a cut at $z=0$, while the upper half shows the magnetic field lines in blue. The rows present snapshots at different times $t=\{0.00, 2.46, 4.93, 7.39, 9.85 \} \ \rm ms$ for the individual simulations. The top row shows the results of the simulation of a single NS solving the Tolman–Oppenheimer–Volkoff (TOV) equations, whereas the second to fourth rows show respectively the results of the simulations ecc0.50, ecc0.55 and ecc0.60 with R2 resolution. Below, the proper distance $d$ for each simulation is plotted with the snapshot times marked by dashed, vertical lines.}
    \label{fig:FirstEncounter}
\end{figure*}

While the results are generally consistent with the predicted expectations, there are small differences in the signals for different resolutions. Especially for the ecc0.50 simulations, there are visible differences on later time scales and we observe a small shift in the signals. However, the lines for the simulations at R1 resolution with and without magnetic field overlap perfectly in every case. Thus, our results show that even for eccentric BNS systems and several encounters before the merger, the magnetic field appears to be dynamically irrelevant before the merger and the GW signal in particular. Also, the $f$-mode oscillations do not seem to be influenced by the inclusion of the magnetic field inside the NSs. 
This is in agreement with previous studies on single, highly magnetized NS, e.g., \cite{Leung:2022mvm,Yip:2023hmo,Yip:2023zza,Yip:2023qkh}. The magnetic field only influences stellar oscillations if the ratio between the magnetic and binding energy is above $10\%$~\cite{Leung:2022mvm}, while Ref.~\cite{Yip:2023qkh} shows that for a maximum magnetic field strength $B_{\rm max} \approx 10^{17}\ \rm G$, i.e., two orders of magnitude higher than in our simulations, the frequency of different oscillation modes is suppressed. 

\subsubsection{Deformation of the Magnetic Field}

Despite having no significant impact on the emitted GW signal for the inspiral, we analyze in this subsection how the magnetic field itself changes during the pre-merger encounters. We restrict the following analysis to one NS of the binary system. In Fig.~\ref{fig:FirstEncounter}, we show three-dimensional snapshots of the magnetic field lines for the first encounter for all systems with R2 resolution. We observe changes in the magnetic field lines mainly near the surface of the star. The lines start to twist, and the magnetic field topology is slightly reconfigured.
It is known that a purely poloidal field structure is unstable to the so-called ``Tayler'' or ``kink'' instability \cite{Wright:1973,Markey:1973}. This has also been shown in numerous numerical simulations, e.g., \cite{Lasky:2011un,Ciolfi:2011xa,Ciolfi:2012en}. The magnetic field inside the NS is rearranged within a few Alfv\'{e}n time scales leading to a mixed configuration of toroidal and poloidal field.

Therefore, we additionally simulated a single NS solving the Tolman-Oppenheimer-Volkoff (TOV) equations for the same EOS, mass, and magnetic field structure with the same resolution as in the BNS simulations. Figure~\ref{fig:FirstEncounter} contains snapshots of the TOV simulation in the top row at the same times as the BNS simulations. This allows us to distinguish changes in the magnetic field structure caused by the encounter from those caused by the simple evolution of the NS itself, i.e., by the unstable configuration.
Indeed, we observe similar changes in the magnetic field configuration inside the star in the case of a single NS. Again, the magnetic field lines near the star's surface begin to oscillate and bend. However, this happens on slightly different time scales. In Fig.~\ref{fig:FirstEncounter}, the first clear changes in the magnetic field structure inside the star can already be seen in the third snapshot at $4.93\ \rm ms$ for the BNS systems, which coincides with the time just before the first encounter. For the TOV star, such features are only visible in later snapshots. Nevertheless, in the last snapshot at about $ 9.85\ \rm ms$, the magnetic field lines are twisted in a similar way for all configurations, the eccentric BNS systems as well as the individual NS.

\begin{figure}[t!]
    \centering
    \includegraphics[width=\linewidth]{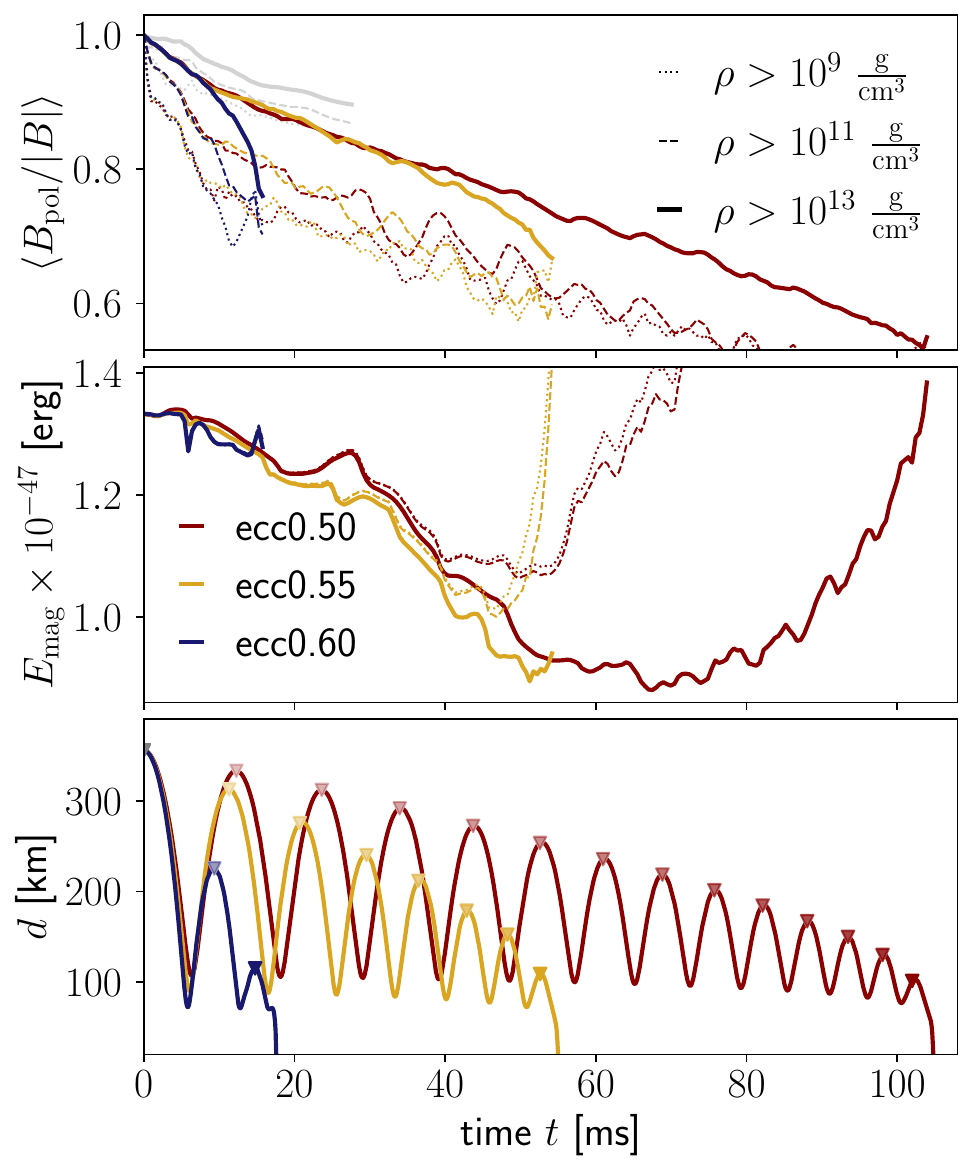}
    \caption{Temporal evolution of the magnetic field inside the stars at R2 resolution. \textit{Top panel}: The poloidal magnetic field component $B_{\rm pol}$ normalized to the field strength $|B|$ averaged over $\rho > \{10^{9},10^{11},10^{13}\} \ \frac{\rm g}{\rm cm^3}$, respectively dotted, dashed, and solid lines. \textit{Middle panel}: The magnetic energy $E_{\rm mag}$ integrated over $\rho > \{10^{9},10^{11},10^{13}\} \ \frac{\rm g}{\rm cm^3}$, respectively dotted, dashed, and solid lines. \textit{Bottom panel}: The evolution of the proper distance $d$. The quantities for the top and middle panel are extracted from refinement level $l=8$. We add in the top panel results of the single NS simulation in gray lines. Also, we mark in the bottom panel the times for which we compute the magnetic spectra in Fig.~\ref{fig:FirstEncounterSpectra} by triangles. }
    \label{fig:FirstEncounter1D}
\end{figure}

For a quantitative assessment of how the magnetic field topology changes during the close encounters before merger, we show in Fig.~\ref{fig:FirstEncounter1D} the decay of the poloidal structure together with the evolution of the magnetic energy within the stars. In particular, we show the evolution of the average poloidal magnetic field component $B_{\rm pol}$, normalized to the field strength $|B|$, and the integrated magnetic energy $E_{\rm mag}$ over different density thresholds. The solid lines for $\rho > 10^{13}\ \frac{\rm g}{\rm cm^3}$ represent the magnetic field entirely inside the NS, whereas the dashed and dotted lines also partially cover the outside of the star. In order to relate the temporal evolution of the magnetic field to the close encounters, the lower panel again shows the proper distance between the two stars. 

As expected, $\left< B_{\rm pol} / |B| \right>$ decreases with time, confirming the decay of the poloidal field structure. While the differences between the ecc0.50, ecc0.55, and ecc0.60 systems are only minor, $\left< B_{\rm pol} / |B| \right>$ decreases faster when the lower density regions outside the star are included. This confirms the earlier observation that the oscillations start at the star's surface and lower density region outside of the star. Closer to the merger, the lines for $\rho > 10^{13}\ \frac{\rm g}{\rm cm^3}$ decrease quite rapidly for each case. We attribute this to the turbulence during the merger winding up the magnetic field causing an increase in the toroidal component.
For comparison, we also include the results of the single NS simulation in gray lines in Fig.~\ref{fig:FirstEncounter1D}. The decay rate for the single star is noticeably smaller compared to the stars in the binary systems, which also explains the later restructuring of the magnetic field observed in Fig.~\ref{fig:FirstEncounter}. 
We find a small dependence between the decay of the poloidal structure and the timings of the close encounters. In particular, the lines for $\rho > 10^{11}\ \frac{\rm g}{\rm cm^3}$ show a clear drop when the stars approach, indicating that instabilities are triggered. This seems to be independent on the eccentricity and the impact parameter, as the strength of the poloidal component are overall of similar order in all BNS systems.

\begin{figure}[t!]
    \centering
    \includegraphics[width=\linewidth]{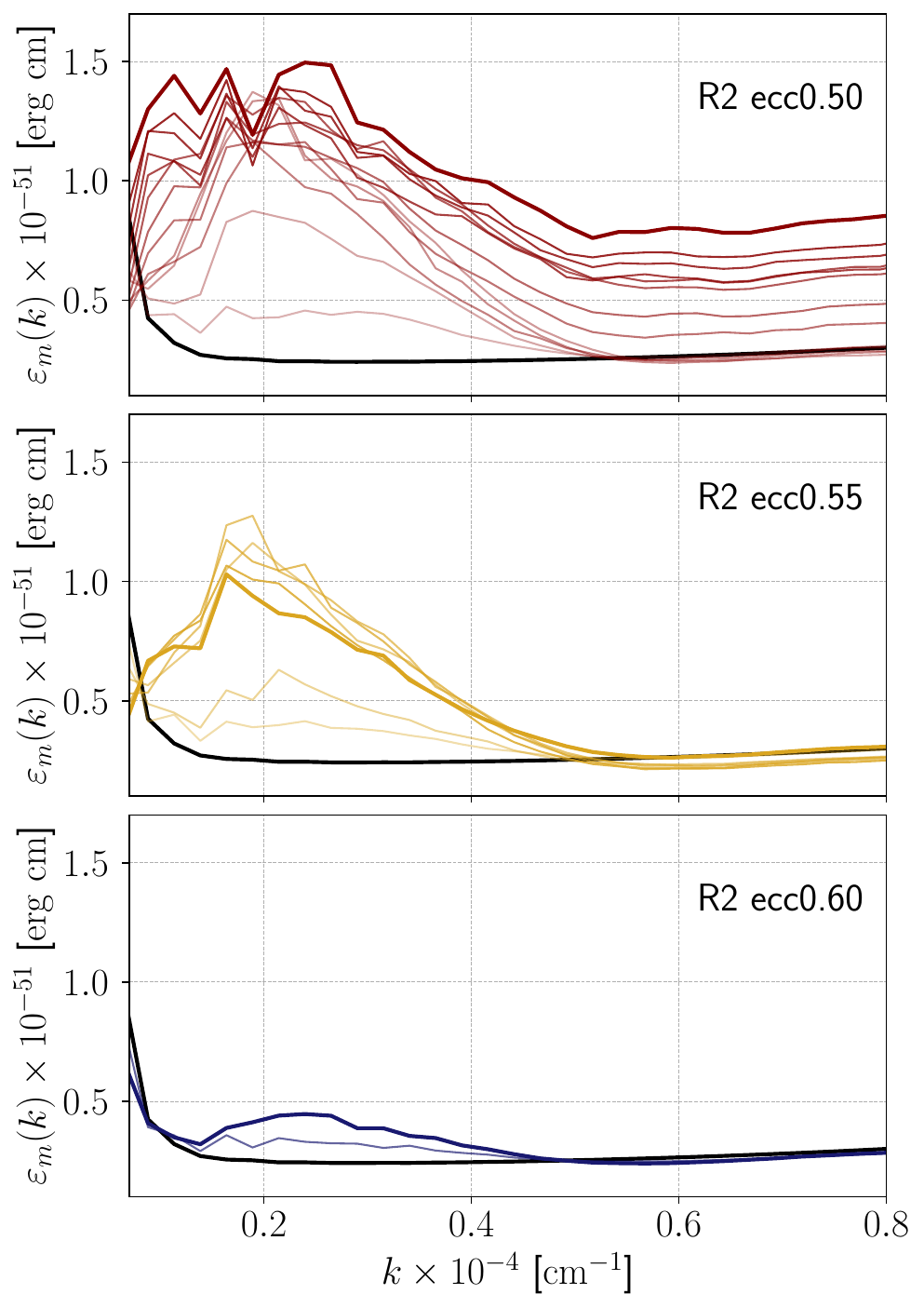}
    \caption{Magnetic energy spectra inside the NS for $\rho<10^{13}\ \frac{\rm g}{\rm cm^3}$ for the ecc0.50, ecc0.55, and ecc0.60 simulations at R2 resolution. We show the spectra for the times at which the stars are in apoapsis, marked in Fig.~\ref{fig:FirstEncounter1D} by triangles in the same color as the corresponding lines. The initial magnetic energy distribution for $t=0\ \rm ms$ is shown as black lines. We compute $\varepsilon_m(k)$ on refinement level $l=8$.}
    \label{fig:FirstEncounterSpectra}
\end{figure}

At the same time, the magnetic energy inside the NS decreases slightly with each close encounter, and it appears that the energy decreases faster at lower eccentricities. After $\sim 65 \ \rm ms$, however, the magnetic energy in the NS increases again in the ecc0.50 simulation. Since the other two systems have already merged at this point, it is unclear whether this feature would also occur in the other systems if they had evolved over a longer time scale. Nevertheless, we note that the changes in the magnetic energy are relatively small and only about $\sim 30 \%$. 

We further analyze how the spectral magnetic energy distribution changes inside the stars after each encounter. For this purpose, we pick the times when the stars are at apoapsis and calculate the magnetic energy spectra as in Eq.~(13) of \cite{Palenzuela:2021gdo}
\begin{equation}
    \varepsilon_m(k) = \frac{L^3 4 \pi}{\left(2 \pi\right)^3 N^6} \left< k^2 \Tilde{b}(k) ^2\right>_k.
\end{equation}
Here, $L$ is the domain size over which the fast Fourier transformation for $\Tilde{b}$ is performed, $N$ the number of points per direction, $k$ the radial wavenumber, and $b^2$ the magnetic energy density. In order to compute the magnetic energy spectra only within the star, we set $b^2$ for $\rho<10^{13}\ \frac{\rm g}{\rm cm^3}$ to zero. We calculate $\varepsilon_m(k)$ at the finest refinement level and show them in Fig.~\ref{fig:FirstEncounterSpectra}. The corresponding times for which the magnetic energy distribution is calculated are also marked in the bottom panel of Fig.~\ref{fig:FirstEncounter1D}. 

Overall, the magnetic energy distribution does not change significantly for wavenumbers above $0.55 \times 10^{-4} \ \rm cm^{-1}$. Only for the ecc0.50 system, the energy increases for these spatial scales after the fifth apoapsis time at $\sim 60\ \rm ms$, when also the total energy inside the star rises (see Fig.~\ref{fig:FirstEncounter1D}). Otherwise, the spectral distribution increases for wavenumbers between $0.1 \times 10^{-4} \ \rm cm^{-1}$ to $0.4 \times 10^{-4} \ \rm cm^{-1}$, corresponding to a length scale of $\lesssim 1\ \rm km$, after each close encounter. This could be attributed to the oscillations visible in the magnetic field lines in Fig.~\ref{fig:FirstEncounter}. Although we note that the differences here are also minor, the shift in the magnetic energy distribution appears to be slightly stronger for the BNS system with smaller eccentricity when comparing the first few encounters.

\subsection{Merger Dynamics}

\subsubsection{Magnetic Field Amplification}

\begin{figure}[t!]
    \centering
    \includegraphics[width=\linewidth]{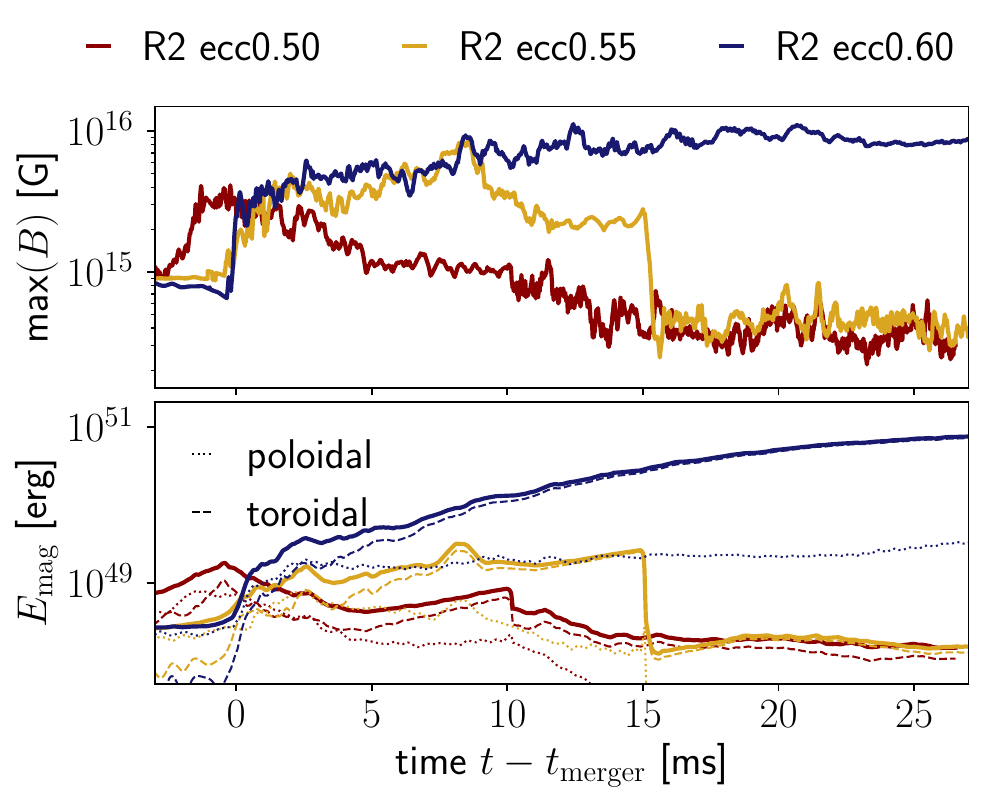}
    \caption{Temporal evolution of maximum magnetic field strength (upper panel) and the magnetic energy (lower panel) for the BNS simulations with R2 resolution. The dashed lines show the toroidal part and dotted lines poloidal part of the magnetic energy. The magnetic energy is extracted from refinement level $l=1$ and the maximum magnetic field strength from $l=8$.}
    \label{fig:BNSmagnetic}
\end{figure}

During the merger, the magnetic energy and field strength rise due to a variety of instabilities, e.g., KHI triggered by the shear layer between the two merging stars or MRI induced by the interaction with the remnant disk. To analyze whether these mechanisms depend on the system's eccentricity, we show in Fig.~\ref{fig:BNSmagnetic} the temporal evolution of the magnetic energy and the maximum of the magnetic field strength with respect to the merger time. Because capturing the instabilities and the corresponding magneto-hydrodynamic turbulence leading to magnetic field amplification require high resolution, we focus in Fig.~\ref{fig:BNSmagnetic} on results with R2 resolution, but note that the amplification for the R1 simulations is significantly lower. In fact, even higher resolutions than our R2 configuration are required to fully capture these instabilities \cite{Kiuchi:2015sga, Kiuchi:2017zzg,Kiuchi:2024lpx}.

As expected, the magnetic field amplifies at the time of merger, reaching values of $\sim 4 \times 10^{15}\rm G$. For the system ecc0.50, the magnetic field strength even rises shortly before merger and already increases during the evolution (as also shown in Fig.~\ref{fig:FirstEncounter1D}). However, the field strength decreases quite rapidly afterward, falling back to the previous values, while it continues to increase to $10^{16}\ \rm G$ in the other two systems. At about $15\ \rm ms$ after the merger, the magnetic field for the ecc0.55 system also drops drastically to values similar to those of the ecc0.50 system. We explain this rapid decrease in both cases by the collapse of the remnant. For the ecc0.50 and ecc0.55 systems, a black hole forms at, respectively, $11 \ \rm ms$ and $15\ \rm ms$ after merger, while there is no black hole for the ecc0.60 system in the simulation time.

The magnetic energy $E_{\rm mag}$ increases for the ecc0.60 system up to $10^{51}\rm erg$. In particular, the toroidal contribution amplifies, while the poloidal part remains more or less constant after the initial amplification during the merger. Here, the poloidal and toroidal components are determined with respect to the coordinate center, which is only approximately correct after merger assuming the remnant is located at the center.
Also for the ecc0.55 system, the magnetic energy increases, but less strongly and only up to $10^{49}\rm erg$. However, for the ecc0.50 system, $E_{\rm mag}$ is already $\sim 10^{49}\rm erg$ before the merger and is not significantly enhanced during the merger itself. After the collapse in the ecc0.50 and ecc0.55 systems, the magnetic energy in both systems drops to similar values of $\sim 10^{48}\rm erg$.

Overall, the magnetic enhancement during the merger seems to be independent of the eccentricity, and all systems reach similar values for the magnetic field strength and energy in the first millisecond after the merger. Nevertheless, we emphasize that to fully capture the KHI triggering the amplification in the early phase of the merger, a higher resolution might be needed to be able to make an accurate statement. Later, at $\sim 10 \ \rm ms$ after the merger, our results indicate to reach higher magnetic energies for higher eccentricities. On these time scales, the toroidal component is dominant and increases due to MRI and magnetic winding.
After the collapse of the remnant in the ecc0.50 and ecc0.55 systems, $E_{\rm mag}$ declines, which seems to be independent of the eccentricity, while in the ecc0.60 system the magnetic energy continues to increase powered by the remnant massive neutron star. 

\subsubsection{Mass Ejection}

\begin{figure}[t!]
    \centering
    \includegraphics[width=\linewidth]{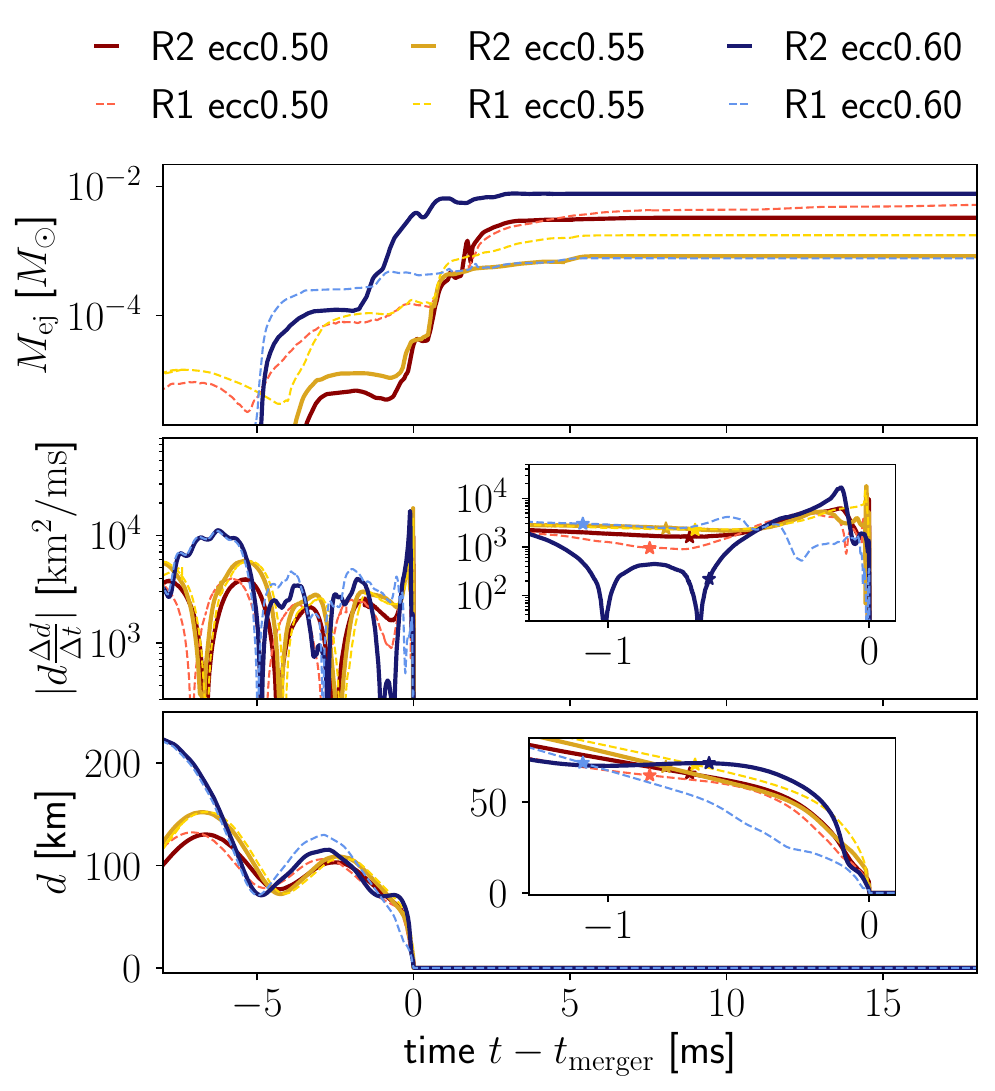}
    \caption{Time evolution of the ejecta mass $M_{\rm ej}$ (top panel) and the proper distance $d$ (bottom panel) around the BNS merger for all simulations with R1 (dashed lines) and R2 (solid lines). Additionally, we show the temporal change in the separation scaled with the distance $|d \frac{\Delta d}{\Delta t}|$ (middle panel) as an estimate on the radial impact velocity. The ejecta mass is extracted from refinement level $l=1$ and the proper distance computed on refinement level $l=8$. The stars mark the time at which the two NSs get in contact (see also Fig.~\ref{fig:BNS_lowres} and Fig.~\ref{fig:BNS_highres}).}
    \label{fig:BNSejecta}
\end{figure}

In general, eccentric BNS systems are expected to eject a larger amount of material than quasi-circular ones \cite{Radice:2016dwd,Papenfort:2018bjk,Chaurasia:2018zhg}. Additionally, material can already be ejected by strong tidal torques at the close encounters before merger \cite{Chaurasia:2018zhg}. We show in Fig.~\ref{fig:BNSejecta} the temporal evolution of the ejected mass for the simulated BNS systems with R1 and R2 resolution around the time of the merger. Following the geodesic criterion for the ejecta, we are considering matter to be unbound if $u_t < -1$ and $v_r >0$. 

In each simulation, a small amount of material is ejected before merger by tidal interaction. Considering the results for each resolution separately, the unbound matter before merger seems to be larger for systems with higher eccentricity. More specifically, at $\sim 2.5\ \rm ms$ before merger, for the simulations with resolution R1 we find an ejected mass of $\gtrsim 10^{-4}\ M_\odot$ for ecc0.60 and $\lesssim 10^{-4}\ M_\odot$ for ecc0.50 and ecc0.55, and for the simulations with R2 resolution $\sim 10^{-4}\ M_\odot$ for ecc0.60 and $\sim 10^{-5}\ M_\odot$ for ecc0.50 and ecc0.55. 
After merger, the R1 simulations yield higher ejecta masses for lower eccentricities, ranging from $8 \times 10^{-4}\ M_\odot$ to $6 \times 10^{-3}\ M_\odot$. This agrees with results from \cite{Chaurasia:2018zhg}, where similar equal-mass BNS systems were studied with SLy EOS but without a magnetic field: The amount of unbound matter decreases with increasing eccentricity. However, this is not the case in the R2 simulations. Here, the system ecc0.60 with $8 \times 10^{-3}\ M_\odot$ ejects the largest amount of material and the system ecc0.55 with $8 \times 10^{-4}\ M_\odot$ the least.

\begin{figure}[t]
    \centering
    \includegraphics[width=\linewidth]{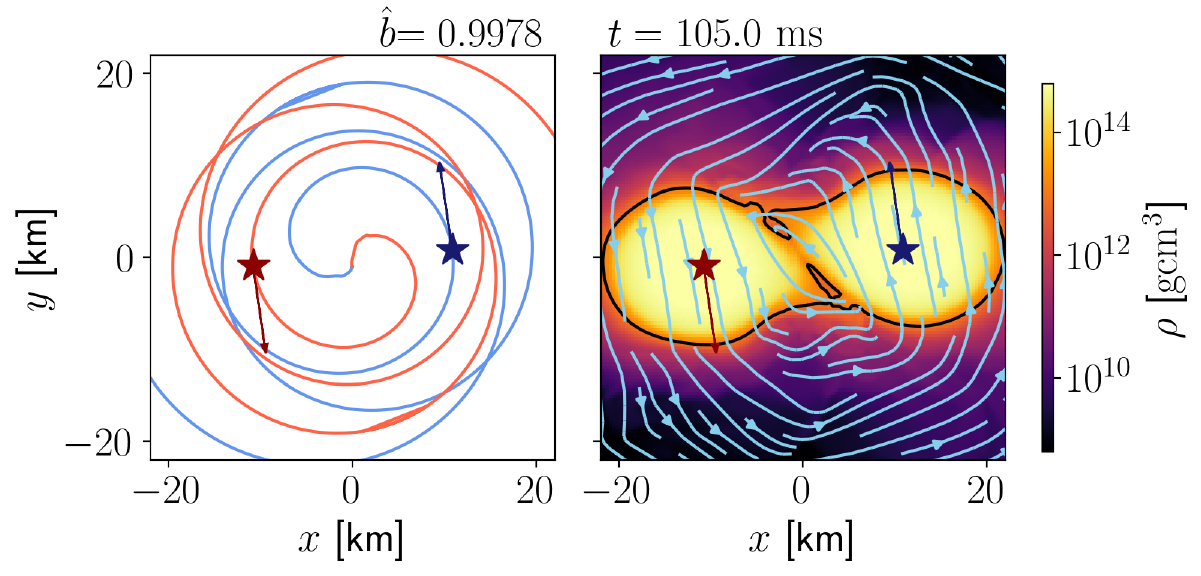}
    {\hfill (a) R1 ecc0.50}
    \includegraphics[width=\linewidth]{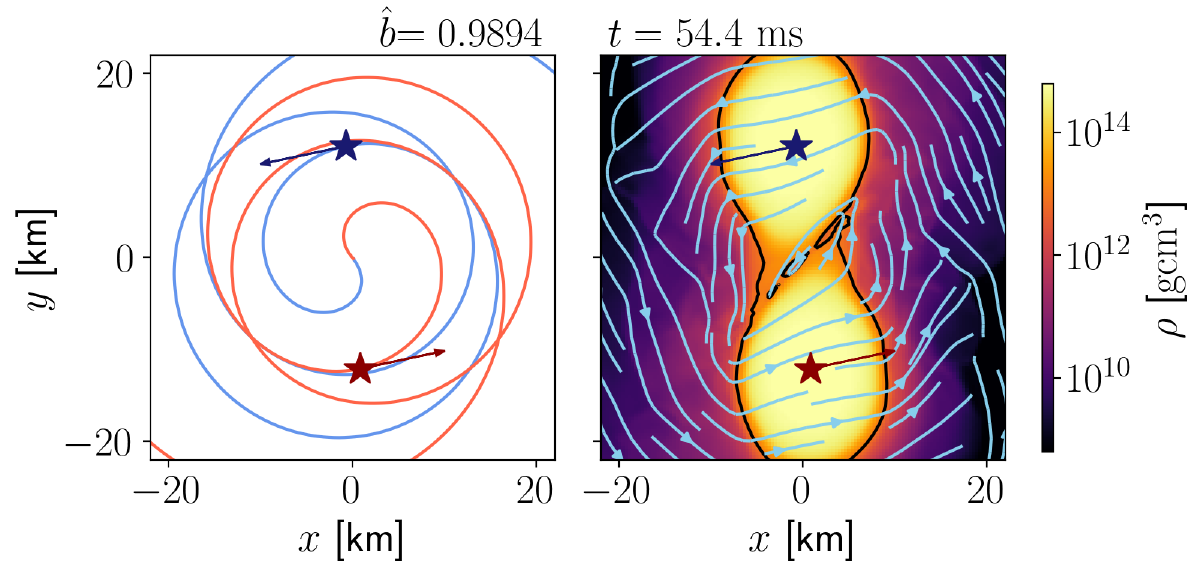}
    {\hfill (b) R1 ecc0.55}
    \includegraphics[width=\linewidth]{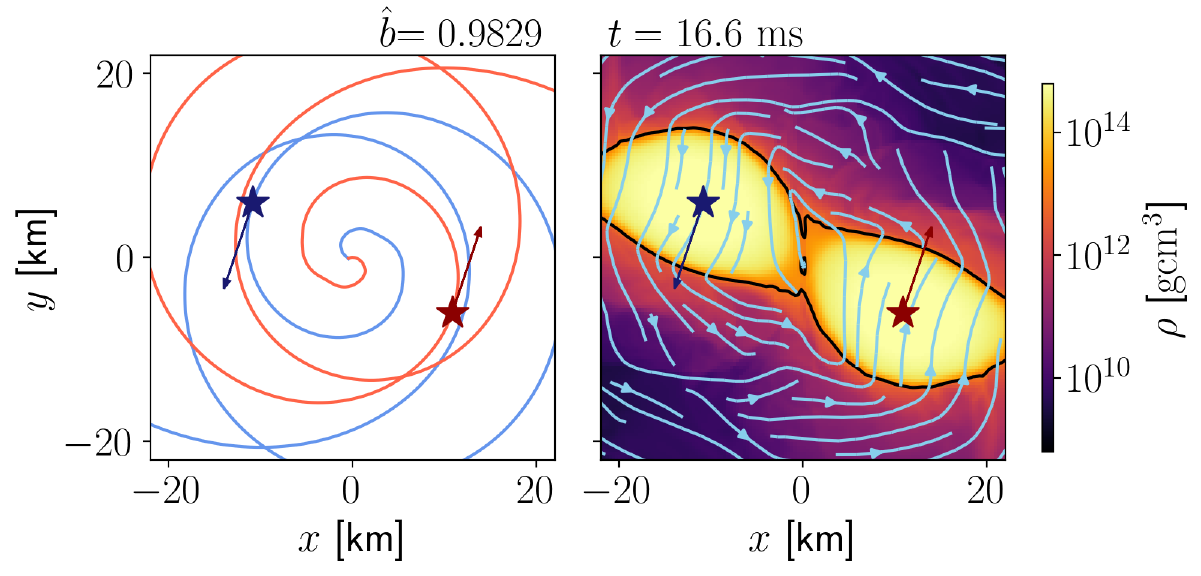}
    {\hfill (c) R1 ecc0.60}
    \caption{Collision of the BNS systems with R1 resolution. For each system, we show on the left the orbital trajectories of the stars' center in blue and red lines, and on the right snapshots in the $x$-$y$ plane of the rest-mass density $\rho$ together with black contour lines for $\rho = 10^{13}\ \frac{\rm g}{\rm cm^3}$ and streamlines of the fluid velocity. The position of the NS centers at the corresponding time is marked by red and blue stars. We add arrows for the velocity based on the star trajectory and provide the corresponding $\hat{b}$ as an indicator for the impact parameter.}
    \label{fig:BNS_lowres}
\end{figure}

To gain a better understanding, we investigate the collision of the BNS systems in more detail. In fact, the time evolution of the proper distance in the bottom panel of Fig.~\ref{fig:BNSejecta} indicates that the actual impact at merger for ecc0.60 is quite different in the two resolutions. With R1, the distance decreases more steadily than with the R2 simulation. To estimate the impact velocity, we compute the temporal change of the separation scaled by the distance $|d \frac{\Delta d}{\Delta t}|$ (middle panel of Fig.~\ref{fig:BNSejecta}). Indeed, $|d \frac{\Delta d}{\Delta t}|$ for the system ecc0.60 increases just before merger for R2 resolution, while for R1 $|d \frac{\Delta d}{\Delta t}|$ is initially higher and slowly decreases.

\begin{figure}[t]
    \centering
    \includegraphics[width=\linewidth]{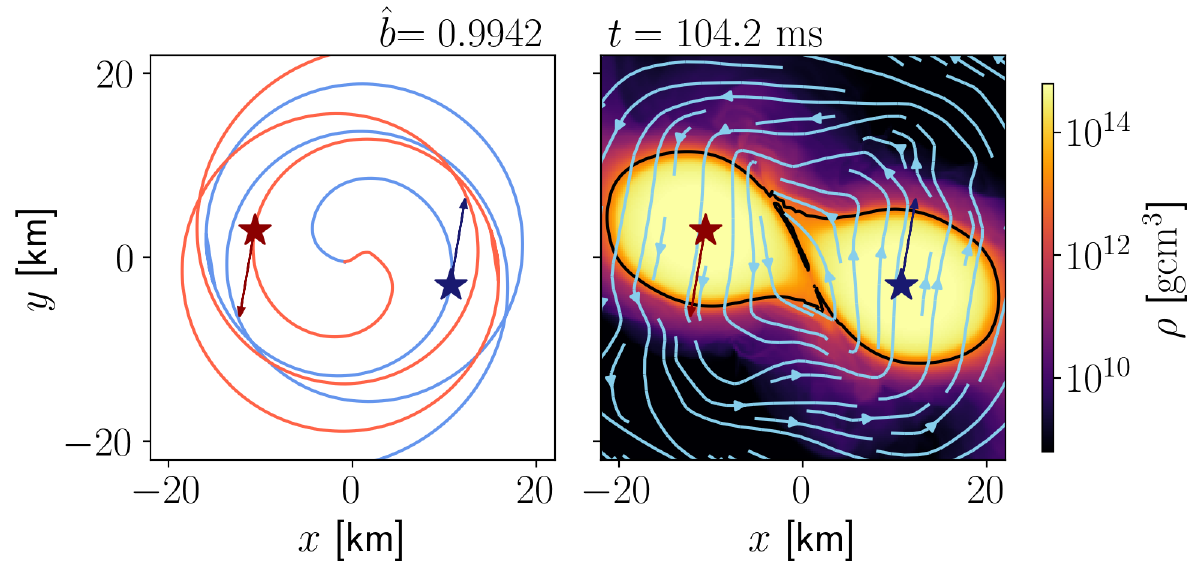}
    {\hfill (a) R2 ecc0.50}
    \includegraphics[width=\linewidth]{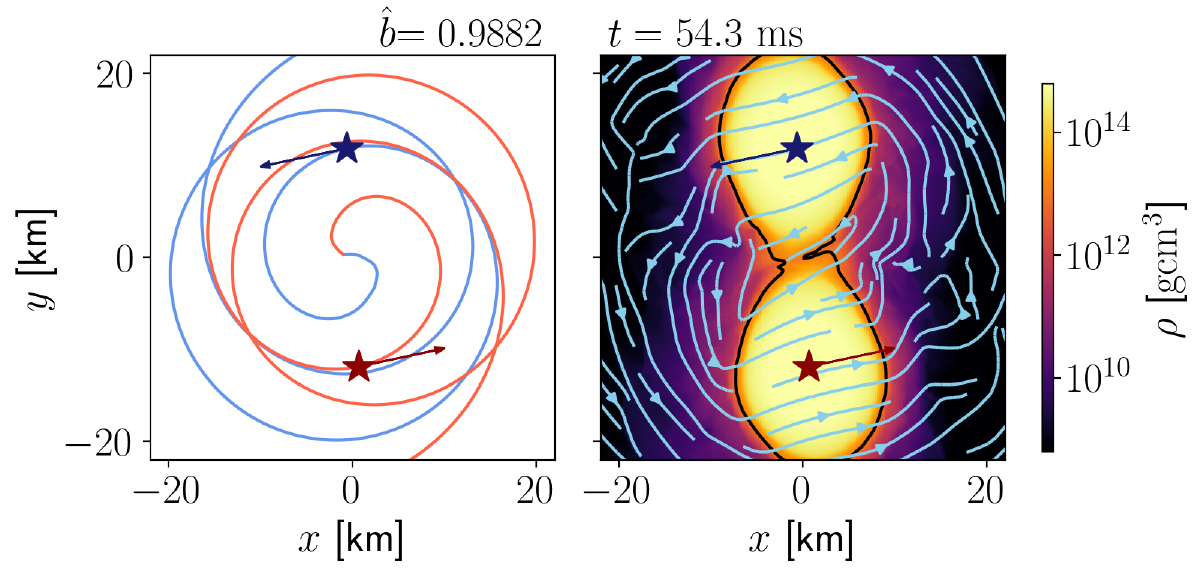}
    {\hfill (b) R2 ecc0.55}
    \includegraphics[width=\linewidth]{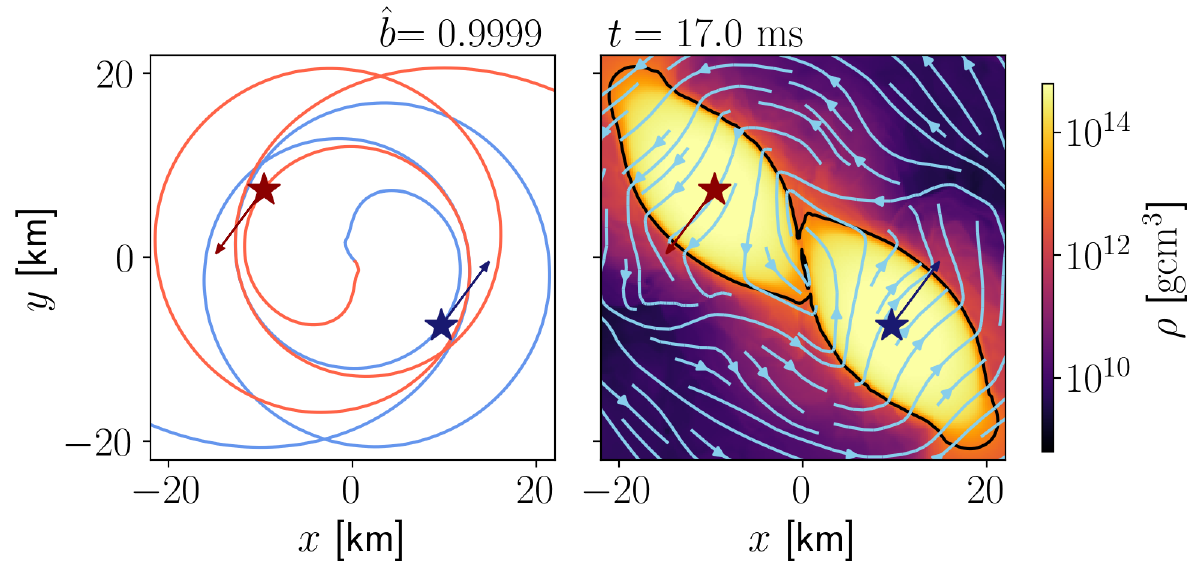}
    {\hfill (c) R2 ecc0.60}
    \caption{Collision of the BNS systems with R2 resolution. For each system, we show on the left the orbital trajectories of the stars in blue and red lines, and on the right snapshots in the $x$-$y$ plane of the rest-mass density $\rho$ together with black contour lines for $\rho = 10^{13}\ \frac{\rm g}{\rm cm^3}$ and streamlines of the fluid velocity. The position of the NS centers at the corresponding time is marked by red and blue stars. We add arrows for the velocity based on the star trajectory and provide the corresponding $\hat{b}$ as an indicator for the impact parameter.}
    \label{fig:BNS_highres}
\end{figure}

To capture the impact when the NSs collide, we try to find out when the two stars get in contact. We define the NS by $\rho > 10^{13}\ \frac{\rm g}{\rm cm^3}$, which is often used to define a NS remnant, e.g., \cite{Shibata:2017xdx,Kiuchi:2019lls,Vincent:2019kor,Radice:2018pdn,Schianchi:2023uky}. The timings when the NS `touch' are marked by stars in Fig.~\ref{fig:BNSejecta} and corresponding snapshots are provided in Fig.~\ref{fig:BNS_lowres} for R1 simulations and Fig.~\ref{fig:BNS_highres} for R2 simulations. We actually find that a lower $|d \frac{\Delta d}{\Delta t}|$ at contact results in a higher ejecta mass. This indicates that, at higher radial impact velocities, less material is ejected for the simulated eccentric BNS systems. 

We present in Fig.~\ref{fig:BNS_lowres} and Fig.~\ref{fig:BNS_highres} snapshots of the BNS systems in the $x$-$y$ plane showing the density profile and streamlines of the fluid velocity together with trajectories of the NS's center ${\bf r}_{\rm A}$ and ${\bf r}_{\rm B}$. From the latter, we compute the corresponding velocity ${\bf v}_{\rm A,B} = \Delta {\bf r}_{\rm A,B} / \Delta t$, which is in good agreement with the fluid velocity at the NS center. In order to obtain an indicator for the impact parameter, we define:
\begin{equation}
    \hat{b}_{\rm A,B} = \frac{\left| {\bf v}_{\rm A,B} \times {\bf r}_{\rm A,B} \right|}{\left| {\bf v}_{\rm A,B}\right| \left| {\bf r}_{\rm A,B} \right|},
\end{equation}
where we have $\hat{b}=\hat{b}_{\rm A}=\hat{b}_{\rm B}$ due to the symmetry of our system. The corresponding $\hat{b}$ is provided in Fig.~\ref{fig:BNS_lowres} for R1 simulations and Fig.~\ref{fig:BNS_highres} for R2 simulations. Specifically, $\hat{b}$ describes whether the stars collide frontally for $\hat{b}\rightarrow 0$ or orthogonally for $\hat{b} \rightarrow 1$. For the simulated BNS merger, $\hat{b}$ is relatively high, i.e., larger than $0.98$, for all systems. 

\begin{figure}[t!]
    \centering
    \includegraphics[width=\linewidth]{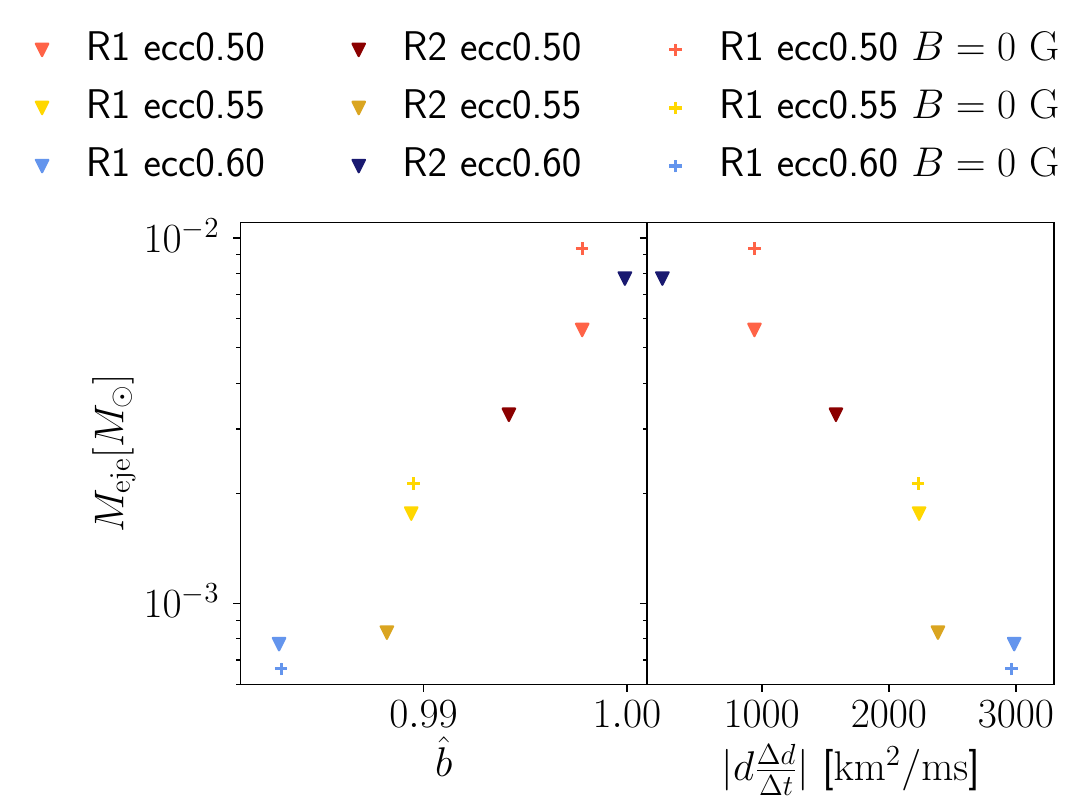}
    \caption{Relation between the ejecta mass $M_{\rm eje}$ and the impact indicator $\hat{b}$ (left panel) and, respectively, $|d \frac{\Delta d}{\Delta t}|$ (right panel) as estimate on the radial impact velocity, for all simulations at R1 and R2 resolution as well as the simulations without magnetic field at R1 resolution.}
    \label{fig:BNScorr}
\end{figure}

The highest $\hat{b}$ is reached for the R2 ecc0.60 simulation with $0.9999$, which also ejects the largest amount of material. The corresponding R1 simulation shows the lowest value with $\hat{b}=0.9828$, but also the lowest ejecta mass. In Fig.~\ref{fig:BNScorr}, we show the respective values for $\hat{b}$ and $|d \frac{\Delta d}{\Delta t}|$ related to the corresponding ejecta masses for all simulated systems, including the simulations without magnetic field. We find clear tendencies that larger $\hat{b}$ leads to larger amount of ejecta, while larger $|d \frac{\Delta d}{\Delta t}|$ leads to less ejecta. This agrees with the results of \cite{East:2012ww} that a lower impact parameter decreases the amount of unbound matter. This result seems to be independent of whether or not magnetic fields are considered.

\subsection{Remnant System}

In most of the performed simulations, the remnant collapses within the simulation time. Only in the R2 ecc0.60 simulation no black hole (BH) is formed.
In Tab.~\ref{tab:remnant} we summarize some key quantities of the remnant system, including the ejecta mass, the remnant BH mass and its spin, obtained at the end of the simulations at $\sim 30\ \rm ms$ after the merger. In addition, we list $\hat{b}$ from Fig.~\ref{fig:BNS_lowres} and \ref{fig:BNS_highres} as impact indicators. We point out that in the ecc0.50 simulations our apparent horizon finder failed. Since we compute the mass $M_{\rm BH}$ and the dimensionless spin parameter $\chi_{\rm BH}$ from the apparent horizon, we cannot provide reliable results for these simulations. We also provide in Tab.~\ref{tab:remnant} the collapse time $t_{\rm col}$, which we define as the time between the merger and the time when an apparent horizon is first found. In the case of the ecc0.50 simulations, we estimate $t_{\rm col}$ using the minimum of the time lapse $\alpha$. 

As discussed above, we find an increase in the ejecta mass for higher $\hat{b}$. Similarly, the collapse time is delayed and the BH mass is smaller when $\hat{b}$ is larger. However, this could also be simply an effect of the larger matter outflow leading to less bound material that can be compressed. In Tab.~\ref{tab:remnant}, we also list the results of the simulations without magnetic field with R1 resolution. For the ecc0.50 and ecc0.55 systems, the ejecta masses with $B=0\ \rm G$ are larger by a factor of $1.67$ and $1.21$, respectively. However, for the ecc0.60 simulation, more ejecta is produced with magnetic field. Furthermore, we find a delayed collapse if the magnetic field is not taken into account. This is probably due to enhanced angular momentum transport when magneto-hydrodynamic effects are considered.\\

\begin{table}[t]
\caption{Properties of the remnant system, from left to right: name, resolution, impact indicator $\hat{b}$, ejecta mass, BH mass, dimensionless BH spin, and collapse time. The quantities are extracted at the end of each simulation at $\sim 30\ \rm ms$ after merger. For the ecc0.50 systems, our apparent horizon finder failed, which is why we have no estimate of the BH mass and spin, and the collapse times (marked with $^*$) are approximated based on the minimum of the lapse $\alpha$. }
\label{tab:remnant}
\begin{tabular}{c c||c|c|c c|c}
\toprule 
name  & res.  & $\hat{b}$ & $M_{\rm eje} [M_{\odot}]$ & $M_{\rm BH} [M_{\odot}]$ & $\chi_{\rm BH}$ & $t_{\rm col} [\rm ms]$  \\ 
\hline \hline
ecc0.50 & R1 &  0.9978 & 0.00559 & $\times$ & $\times$ &  12.34$^*$ \\
ecc0.55 & R1 &  0.9894 & 0.00176 &  2.5005 & 0.6522 &  14.87 \\
ecc0.60 & R1 &  0.9829 & 0.00077 &  2.4496 & 0.6268 &  19.65 \\
ecc0.50 & R2 &  0.9942 & 0.00328 & $\times$ & $\times$ &  11.57$^*$ \\
ecc0.55 & R2 &  0.9882 & 0.00083 &  2.5280 & 0.6702 &  15.34 \\
ecc0.60 & R2 &  0.9999 & 0.00773 &   --    &  --    &  --    \\
\hline
$B=0\ \rm G$ & & & & & & \\
ecc0.50 & R1 & 0.9978 & 0.00933 & $\times$ & $\times$ &  15.50$^*$ \\
ecc0.55 & R1 & 0.9895 & 0.00213 &  2.5089 & 0.6503 &  15.30 \\
ecc0.60 & R1 & 0.9830 & 0.00066 &  2.4596 & 0.6404 &  23.31 \\
\bottomrule
\end{tabular}
\end{table}

\begin{figure*}[t]
    \centering
    \includegraphics[width=\linewidth]{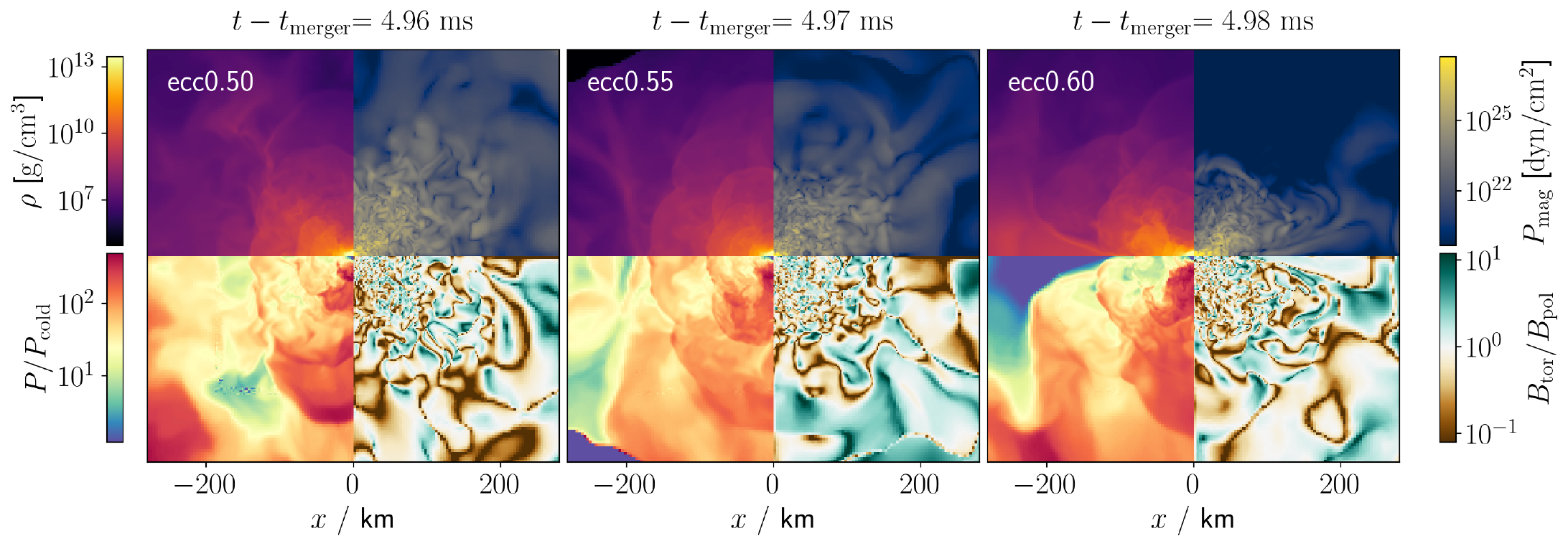}
    \includegraphics[width=\linewidth]{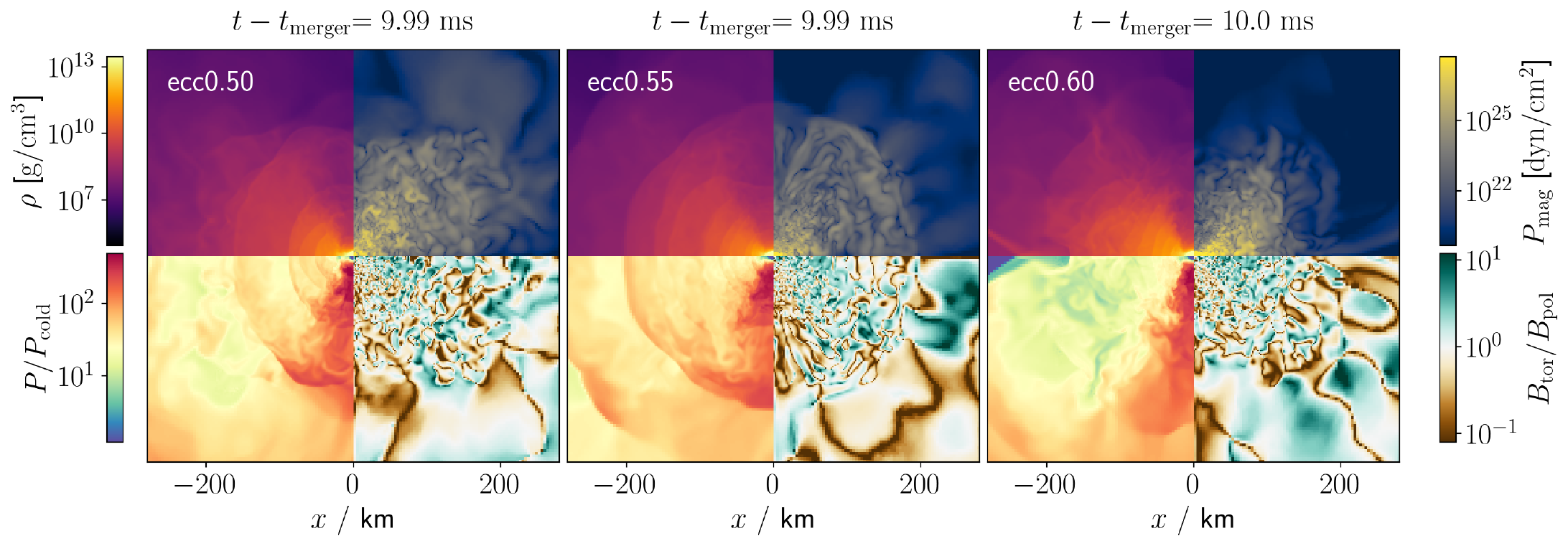}
    \includegraphics[width=\linewidth]{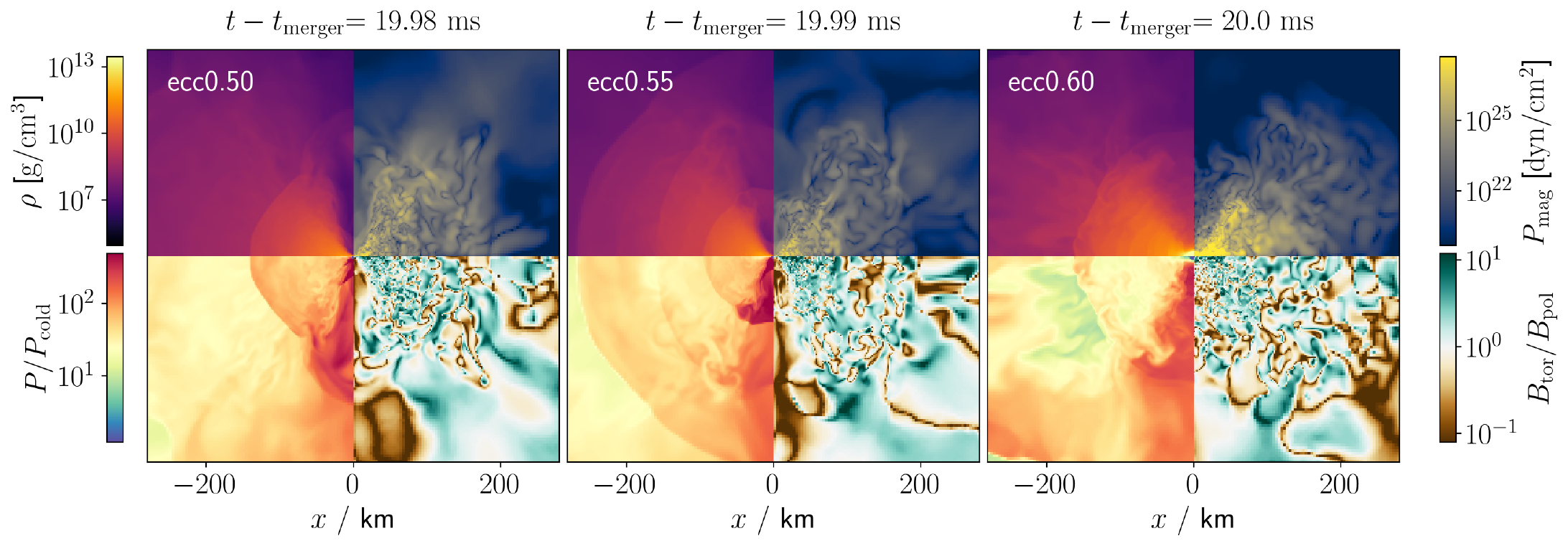}
    \caption{Snapshots of the ejecta and remnant for the BNS simulations at R2 resolution in the $x$-$z$ plane. Each row represents a specific time after the merger, i.e., $t-t_{\rm merger} \approx \{5, 10, 20\} \ \rm ms$, and each column a BNS system with different input eccentricity. The individual panels show the rest-mass density (top left corner), the magnetic pressure (top right corner), the ratio between the pressure and its zero-temperature contribution as an entropy indicator (bottom left corner), and the ratio between the toroidal and poloidal components of the magnetic field (bottom right corner).}
    \label{fig:BNSremnant}
\end{figure*}

For a qualitative overview of the remnant and matter outflow, we present in Fig.~\ref{fig:BNSremnant} snapshots at about $5 \ \rm ms$, $10 \ \rm ms$, and $20  \ \rm ms$ after merger in the $x$-$z$ plane. In particular, we show the rest-mass density and the magnetic pressure, but also the ratio between the pressure and its zero-temperature contribution as an entropy indicator and the ratio between the toroidal and poloidal components of the magnetic field for simulations with R2 resolution.

As expected, magnetic pressure is highest in the disk, where we have a rather toroidal configuration. The highest values are reached in the ecc0.60 system with about $5 \times 10^{27}\ \rm dyn/cm^{2}$. This agrees with the finding that the magnetic field increases the most in the R2 ecc0.60 simulation (see Fig.~\ref{fig:BNSmagnetic}). However, we observe a stronger magnetization of the ambient medium in ecc0.50 and ecc0.55. We explain this by longer simulation times. As a result, there has been more time for the advection of the magnetic field to transport magnetic pressure to the outer regions. In the last snapshot at $\sim 20  \ \rm ms$ after merger, there is a funnel above the BH remnant for ecc0.50 and ecc0.55. Here, the magnetic field is more poloidal and the magnetic pressure is slightly lower than in the disk.

Based on the entropy indicator $P/P_{\rm cold}$, we can identify which parts of the ejecta have been shocked. If material has been ejected by shock heating, the thermal component of the pressure is expected to be high and therefore $P/P_{\rm cold}$ is large. On the other hand, if matter was unbound by torque, we would expect a low entropy indicator. Accordingly, we would assume high electron fraction for high $P/P_{\rm cold}$ and low electron fraction for low $P/P_{\rm cold}$.
In all systems we find relatively high values of $\gtrsim 10^2$ in the polar region. For ecc0.60, the entropy indicator in the orbital plane is considerably lower than for the other two systems, of the order $10^{-1}$ in the first snapshot at $\sim 5 \ \rm ms$ after the merger. As mentioned above, the ejecta mass is also largest for ecc0.60 and the system has the highest $\hat{b}$ with a rather orthogonal collision. Together with the low entropy indicator, we identify here the tidal disruption as dominant mechanism for matter ejection.
We therefore expect this system to contain a larger amount of low-electron material. Consequently, the associated kilonova should be overall redder, especially for an observer in the equatorial plane, compared to systems with lower eccentricities and impact indicators.

\section{Summary}
\label{sec:Conclusions}

In this work, we investigate eccentric BNS mergers using ideal GRMHD simulations. In total, we consider three different systems with different input eccentricities, $e=0.50$, $e=0.55$, and $e=0.60$, at two resolutions. To the best of our knowledge, these are the first numerical-relativity simulations of BNS systems with high eccentricities that include magneto-hydrodynamic effects. Therefore, we focus our study on magnetic field-related features, as well as on how the inclusion of the magnetic field affects the overall dynamics. We summarize our results below.

Keeping the initial separation between the two stars fixed, the number of simulated orbits until merger varies quite drastically for each system, ranging from $4.2$ for $e=0.60$ to $18.3$ for $e=0.50$. 
In each system we simulate multiple close encounters before the merger, which produce the expected bursts of GW radiation. A prominent feature in the GW signal of eccentric BNS systems is the superimposed signal of $f$-mode oscillations that are excited in the stars by tidal interaction during the close encounters. Overall, we find good agreement between our results and the predicted $f$-mode frequency of $1.76\ \rm Hz$ for a $1.35\ M_\odot$ star using the SLy EOS \cite{Chan:2014kua}. 
Since we want to investigate whether close encounter could trigger magneto-hydrodynamic instabilities that cause magnetic fields to be relevant even before the merger, we compare our results with simulations in which the magnetic field is set to zero. However, no significant difference is observed, indicating that in eccentric BNS systems as well as in quasi-circular ones, magnet fields are not relevant for the overall dynamics in the inspiral phase. Larger magnetic energies with maximum field strength $> 10^{17}\ \rm G$ could lead to differences in the excited stellar oscillations \cite{Leung:2022mvm,Yip:2023qkh}, but those would be several orders of magnitude stronger than in observed pulsars in binary systems \cite{Lorimer:2008se}.

Furthermore, we investigate how the magnetic field inside the NSs modifies during the pre-merger encounters and detect a change in geometry. More specifically, we observe a decay of the initial poloidal magnetic field structure. It is well known that a poloidal structure is unstable and subject to kink instability, leading to a rearrangement of the magnetic field structure, even for single isolated stars. Nevertheless, we find that the poloidal structure in our simulated BNS systems decays faster than in an isolated single star. 
We note that possible changes in the magnetosphere and their interactions are not considered in this work due to the limitations of the ideal GRMHD approximation.

The amplification of the magnetic field during the merger seems to be independent of the eccentricity and reaches similar values for the maximum field strength. We focus here on results obtained with the higher resolution simulation. However, a higher resolution might be needed to fully capture the KHI, causing magnetic field to increase during the merger. On longer timescales, the magnetic energy for the system with $e=0.60$ continues to grow, especially the toroidal component, driven by magneto-hydrodynamic interactions with the massive neutron star remnant. In the other two cases with lower eccentricity, the remnant collapses and the magnetic energy drops to similar values.

The mass ejection, on the other hand, depends strongly on the eccentricity or rather on the impact parameter. We define an indicator for the impact parameter $\hat{b}$ when the two stars come into contact and find a correlation with the amount of matter ejected. A higher impact parameter leads to a larger ejecta mass, agreeing with results found in \cite{Radice:2016dwd,Papenfort:2018bjk,Chaurasia:2018zhg} for eccentric BNS systems neglecting the magnetic field contribution. The highest $\hat{b}$ of $0.9999$, corresponding to an almost orthogonal collision, occurs for the BNS simulation with $e=0.60$ at high resolution. This BNS merger produces the highest ejecta mass, with a significant part formed by tidal disruption. We also observe generally later collapse times for higher $\hat{b}$.
However, we find no clear trend in the ejecta mass whether the magnetic field is considered or not. In the two simulations with lower eccentricity, a larger amount of matter is ejected when the magnetic field is neglected, while in the system with $e=0.6$ more ejecta is found when the effects of the magnetic field are included. Yet, we find a delayed collapse in all systems when the magnetic field is neglected and assume the reason to be an enhanced angular momentum transport by magneto-hydrodynamic processes.

In conclusion, our results show that magnetic fields have no significant influence on the dynamics of the inspiral phase for the examples considered, while their structure changes slightly before merger. The amplification during the merger seems to be independent of the eccentricity. At the same time, the mass ejection depends strongly on the impact parameter, with higher values leading to more ejecta. 
We note that these results are limited by the ideal GRMHD approximation and our finite resolution. To account for effects of interacting magnetospheres, resistive GRMHD simulations would be required and a higher resolution might be needed to fully capture magneto-hydrodynamic instabilities during the merger.

\section*{Acknowledgements}

TD and AN acknowledge support from the Deutsche Forschungsgemeinschaft, DFG, project number 504148597 (DI 2553/7). Furthermore, TD acknowledges funding from the EU Horizon under ERC Starting Grant, no.\ SMArt-101076369. BB acknowledges support from the Deutsche Forschungsgemeinschaft (DFG) under Grant No.\ 406116891 within the Research Training Group RTG 2522/1.
The simulations with \bam\ were performed on the national supercomputer HPE Apollo Hawk at the High Performance Computing (HPC) Center Stuttgart (HLRS) under the grant number GWanalysis/44189, on the GCS Supercomputer SuperMUC\_NG at the Leibniz Supercomputing Centre (LRZ) [project pn29ba], on the HPC systems Lise/Emmy of the North German Supercomputing Alliance (HLRN) [project bbp00049], and on the DFG-funded research cluster jarvis at the University of Potdam (INST 336/173-1; project number: 502227537).

\appendix

\bibliography{ref.bib}

\end{document}